\documentclass[referee]{mn2e}
\usepackage{color}
\usepackage{graphicx}
\usepackage{dcolumn}
\usepackage{bm}

\begin{document}

\newcommand{\myspace}{1.0}
\renewcommand{\baselinestretch}{\myspace}
\Large\normalsize

\title[Density--potential pairs for stellar systems]
      {Density--potential pairs for spherical stellar systems with 
       S\'ersic light--profiles and (optional) power--law cores}
\author[Terzi\'c \& Graham]
       {Bal{\v s}a Terzi\'c$^{1,2}$ and Alister W.~Graham$^3$\\ 
 $^1$Department of Physics, Northern Illinois University,  DeKalb, IL 60115, USA\\
 $^2$Corresponding Author: bterzic@nicadd.niu.edu\\
 $^3$Research School of Astronomy and Astrophysics, Australian National University\\ 
 Private Bag, Weston Creek PO, ACT 2611, Canberra, Australia}

\date{Received 2005 April 13; Accepted 2005 December 31}

\pubyear{2004} \volume{000} 
\pagerange{\pageref{firstpage}--\pageref{lastpage}}

\input{psfig}

\maketitle
\label{firstpage}

\begin{abstract}

Popular models for describing the luminosity--density profiles of
dynamically hot stellar systems (e.g., Jaffe, Hernquist, Dehnen) were
constructed with the desire to match the deprojected form of an
$R^{1/4}$ light--profile.  Real galaxies, however, are now known to
have a range of different light--profile shapes that scale with mass.
Consequently, although highly useful, the above models have implicit
limitations, and this is illustrated here through their application to
a number of real galaxy density profiles.  On the other hand, the
analytical density profile given by Prugniel \& Simien (1997) closely
matches the deprojected form of S\'ersic $R^{1/n}$ light--profiles ---
including deprojected exponential light--profiles.  It is thus
applicable for describing bulges in spiral galaxies, dwarf elliptical
galaxies, and both ordinary and giant elliptical galaxies.  Here we
provide simple equations, in terms of elementary and special
functions, for the gravitational potential and force associated with
this density profile.  Furthermore, to match galaxies with partially
depleted cores, and better explore the supermassive black hole /
galaxy connection, we have added a power--law core to this density
profile and derived similar expressions for the potential and force of
this hybrid profile.  Expressions for the mass and velocity
dispersion, assuming isotropy, are also given.  These spherical models
may also prove appropriate for describing the dark matter distribution
in halos formed
from $\Lambda$CDM cosmological simulations.

\end{abstract}


\section{Introduction}

Both elliptical galaxies and the bulges of disk galaxies, hereafter
collectively referred to as ``bulges'', possess a range of
light--profile ``shapes'' that are well described by S\'ersic's (1963,
1968) $R^{1/n}$ model (e.g., Caon, Capaccioli \& D'Onofrio 1993;
Young \& Currie 1994; 
Graham et al.\ 1996; Graham 2001; Balcells et al.\ 2003).  This model
is a generalisation of de Vaucouleurs' (1948, 1959) $R^{1/4}$ model
which is known to be only appropriate for a subset of elliptical
galaxies having $M_B\sim -21$ mag (e.g., Kormendy \& Djorgovski 1989;
Graham \& Guzm\'an 2003).  The $R^{1/4}$ model's limitation lies in
the fact that it has only two parameters: a radial scale and a surface
brightness scale.  The actual curvature, or ``shape'', of every
$R^{1/4}$ model is the same.

This restriction has carried over into computer simulations of
``bulges''.  This is because the popular models that are used for
describing the luminosity--density profiles of bulges, such as
those from Jaffe (1983), Hernquist (1990), and Dehnen (1993, see also
Tremaine et al.\ 1994),
%
%
were created in order to reproduce an $R^{1/4}$ light--profile when
projected.  All three of these density models have exactly the same
outer profile slope, 
declining with radius as $r^{-4}$.  Consequently, these models are
limited in their ability to a) simulate the range of observed galaxy
structures and b) quantify the evolution of these structures.

Over the past two decades it has also become apparent that the most
luminous ($M_B < -20.5$ mag) elliptical galaxies have
partially--depleted stellar--cores (Kormendy 1985; Lauer 1985;
Ferrarese et al.\ 1994; Lauer et al.\ 1995; Gebhardt et al.\ 1996).
On the other hand, the less luminous elliptical galaxies, sometimes
referred to as ``power--law'' galaxies, have continuously curving
S\'ersic profiles that continue all the way in to the resolution limit
of \textit{HST} images (Trujillo et al.\ 2004).  A promising
explanation, albeit not the only one proffered, for the depleted cores
is that the giant galaxies formed from the dissipationless merger of
two or more fainter elliptical galaxies.  The subsequent gravitational
ejection of stars by the inwardly spiralling supermassive black holes
(SMBHs) --- from the progenitor galaxies -- scours out the core of the
new galaxy (e.g., Begelman, Blandford \& Rees 1980; Makino \&
Ebisuzaki 1996; Faber et al.\ 1997; Milosavljevi\'c \& Merritt 2001,
2005; Graham 2004).
Furthermore, fundamental connections have been found between the SMBH
mass and a bulge's: {\it (i)} magnitude (Magorrian et al.\ 1998;
McLure \& Dunlop 2002; Erwin, Graham \& Caon 2002), {\it (ii)}
light--profile shape (Graham et al.\ 2001, 2003a) and {\it (iii)}
kinematics (Ferrarese \& Merritt 2000, Gebhardt et al.\ 2000).

Given the above connections, 
it is obviously important to have models that are able to unite
properly the domain of the black hole with the rest of the galaxy.  In
this paper we present a modification of Prugniel \& Simien's (1997)
density model that already matches the observed range of `outer' profile
shapes; our modification allows one to additionally model partially
depleted cores.
Moreover, and importantly, we also derive exact expressions for the
potential and force in terms of (fast--to--compute) elementary
functions, making it possible for simulations to explore the influence
of cores and differing profile shapes.  By setting the size of the
partially depleted core to zero, the equations for the potential and
force are applicable to the original Prugniel--Simien density profile, 
for which no previous expressions existed.


In subsection 2.1 we introduce the density model, complete with
power--law core, while subsection 2.2 provides the equations for the
potential and force.  Expressions for the projection of the density
model can be found in subsection 2.3.  In section 3 we fit a number of
popular density models to the luminosity--density profiles of real
galaxies and compare the results with the fit from the
Prugniel--Simien density model and our modification of this model.  A
discussion of relevant issues is given in section 4, and section 5
provides a summary of the main results.  Appendix A provides the
derivation for various expressions pertaining to the velocity
structure of the density profiles, including the circular velocity,
spatial and line--of--sight velocity dispersion.  Appendix B 
provides the derivation of the equations given in section 2.

\section{The Model}

\begin{figure}
\begin{center}
\includegraphics[height=3in]{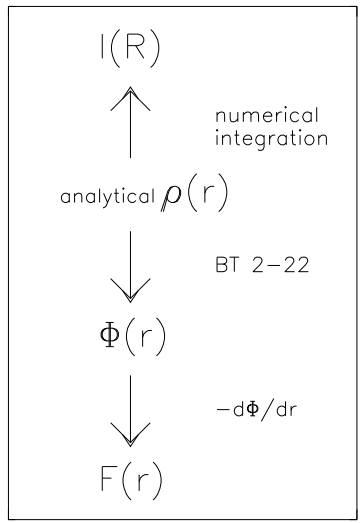}
\caption{Flow chart for the density model: the projection of the 
analytically prescribed density is computed numerically, while the 
potential and force are obtained {\it exactly}.
BT 2--22 refers to Binney \& Tremaine (1987, their equation 2--22).
}
\label{fig1}
\end{center}
\end{figure}

In this section we introduce a family of expressions associated with
the spatial density profiles of spherical stellar systems having
S\'ersic-like profiles with optional power--law cores.  We derive {\it
exact} expressions for the potential and force and outline how the
corresponding surface brightness can be computed numerically (see
Fig.\ref{fig1}).  With computer processing speed in mind, in addition
to the pressing need for more flexible galaxy models, the equations
consist only of analytical terms and simple elementary functions.

For the sake of parameter identification, we first introduce 
S\'ersic's (1963, 1968) model for describing the projected, 
radial intensity profiles of galaxies, such that 
\begin{equation} \label{EqSer}
I(R)=I_0 {\rm e}^{-b(R/R_{\rm e})^{1/n}},
\end{equation}
where $I_0$ is the central intensity and $R_{\rm e}$ is the (projected)
effective half--light radius.  The parameter $n$ describes the
curvature of the profile (see, e.g., Ciotti 1991).  The term $b$ is
not a parameter, but instead a function of $n$ and chosen to ensure
$R_{\rm e}$ contains half the (projected) galaxy light.  It is obtained by
solving the equation $\Gamma(2n) = 2 \ \gamma(2n,b)$, where
\begin{equation} \label{eqgam}
\Gamma(a) = \int_0^{\infty}{\rm e}^{-t}t^{a-1} {\rm d}t 
\hskip20pt {\rm and} \hskip20pt
\gamma(a,x) = \int_0^{x}{\rm e}^{-t}t^{a-1} {\rm d}t, \hskip10pt  a>0 
\end{equation}
are the complete and incomplete gamma functions, respectively. 
Although we have chosen to numerically compute the exact value for $b$
because fast codes are available for the computation of gamma
functions, a good approximation of $b$ for $0.5<n<10$ is
$2n-1/3+0.009876/n$ (Prugniel \& Simien 1997, see MacArthur, 
Courteau, \& Holtzman 2003 for smaller values of $n$). 
Using the substitution $x=b(R/R_{\rm e})^{1/n}$, 
the total luminosity from equation~(\ref{EqSer}) is given as 
\begin{equation} \label{Sertot} 
L_{\rm tot} = \int_0^{\infty} I(R^{\prime})2\pi R^{\prime} {\rm d}R^{\prime} 
            = 2 \pi I_0 R_{\rm e}^{2} n b^{-2n} \Gamma (2n). 
\end{equation} 

For a sample of 250 dwarf elliptical and ordinary elliptical galaxies
spanning $-13 > M_B > -23$ mag, Graham \& Guzm\'an (2003) have shown
how profile shape and surface brightness vary with galaxy magnitude.
They found $M_B = -9.4\log(n) - 14.3$, and $M_B = (2/3)\mu_0 - 29.5$ mag ---
until the presence of cores in galaxies brighter than $\sim$-20.5 
$B$--mag.  The latter phenomenon makes it more appropriate to use $\mu_{\rm e}$, 
the surface brightness at $R_{\rm e}$, rather than the
central surface brightness $\mu_0 = -2.5\log(I_0)$.  Doing so gives
the expression $M_B =(2/3)(\mu_{\rm e} - 1.086b) -29.5$ mag.
Typical values are given Table~\ref{Tabn}. 

A compendium of expressions related to S\'ersic's $R^{1/n}$ model can
be found in Graham \& Driver (2005).

\begin{table}
\caption{Typical galaxy masses and central surface brightnesses $\mu_0
= -2.5\log(I_0)$ (using the inward extrapolation of the outer S\'ersic
profile in the case of ``core'' galaxies) associated with a range of
S\'ersic indices $n$.  The mass estimates have come from the absolute
$B$--band magnitudes in Graham \& Guzm\'an (2003) using
$M_{B,\sun}$=5.47 (Cox 2000) and $M/L_B$=5.31 (Worthey 1994, assuming
a 12 Gyr population with Fe/H=0).  The numbers are only indicative,
with intrinsic variance playing a role amongst real galaxies.  }
\label{Tabn}
\begin{tabular}{lrc}
Mass &  $n$  &  $\mu_{0,B}$ \\
$M_{\sun}$ &   &  mag arcsec$^{-2}$ \\
\hline
$\sim 10^7$   &  0.5 &  $\sim$27 \\
$\sim 10^8$   &    1   &  $\sim$25 \\
$\sim 10^9$   &    2   &  $\sim$20 \\
$\sim 10^{11}$  &    4   &  $\sim$14 \\
$\sim 10^{12}$  &   10  &  $\sim$9  \\
\hline
\end{tabular}
\end{table}

\subsection{Density} 

Generalising an expression from Mellier \& Mathez (1987) that
approximated the spatial, i.e.\ not projected, 
density profile of the $R^{1/4}$ model, Prugniel \&
Simien (1997) provide an analytical approximation to the density
profile of the $R^{1/n}$ model.  Lima Neto, Gerbal \& M\'arquez
(1999) showed that this spherical model is accurate to better than 5\% over the
radial range $10^{-2}$--$10^{3} R_{\rm e}$.  An even more accurate
approximation for both spherical {\it and} triaxial stellar systems with
$R^{1/n}$ light--profiles was developed by Trujillo et al.\ (2002), with an
accuracy better than 0.1\%.  However, although analytical, the latter
expression is not particularly simple.  Impressively, an exact
solution to the deprojection of the $R^{1/n}$ model, which obviously
includes the $R^{1/4}$ case, was given in terms of Meijer G functions
by Mazure \& Capelato (2002); but again these equations are somewhat
complicated\footnote{Exact analytical expressions for the mass,
gravitational potential, total energy and the central velocity
dispersion are also presented in Mazure \& Capelato (2002).  Exact
numerical expressions are given in Ciotti (1991).}.  Moreover, 
although these models can properly treat the range of 
outer profile shapes observed in real galaxies, they can not
additionally allow for the presence of partially depleted cores.  In
fact, we are unaware of any density model capable of simultaneously
describing the range of structure observed in both the inner and outer
regime of galaxies with ``cores''.

\begin{figure*}
\begin{center}
\begin{minipage}{168mm}
\includegraphics[width=168mm]{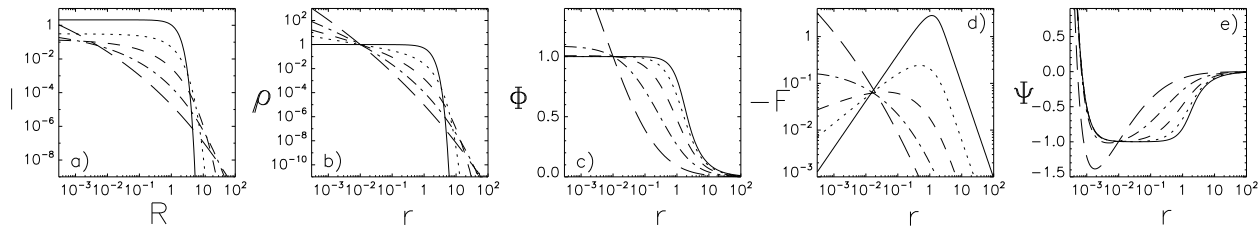}
\vskip30pt
\caption{Core--less galaxies. 
Panel a) projected intensity $I$ (exact numerical solution); 
panel c) potential (equation~\ref{pot_cS2} divided by $\Phi(0.01R_{\rm e})$); 
panel d) force (equation~\ref{force_cS2}); and 
panel e) pseudo potential (equation~\ref{pseudo} 
divided by $|\Phi(0.01R_{\rm e})|$ with $h^2=10^{-6}|\Phi(0.01R_{\rm e})|$) 
associated with the Prugniel--Simien density profiles (equation~\ref{r2_app}) 
shown in panel b) for varying values of the profile shape 
$n$: $n=0.5$ (solid lines), $n=1$ (dotted), $n=2$ (dashed), 
$n=4$ (dash--dot), $n=10$ (double--dash).
The radius and density are normalised such that $R_{\rm e}=1$ 
and $\rho(0.01R_{\rm e})=1$. 
}
\label{ldpf_PS}
\end{minipage}
\end{center}
\end{figure*}

The 3--parameter ($\rho_0, R_{\rm e}, n$) density profile of Prugniel \&
Simien (1997; their equation~B6) is relatively simple and can be written
as
\begin{equation} \label{r2_app}
\begin{array}{ll}
  \rho (r) = \rho_0 \left({r\over R_{\rm e}}\right)^{-p}
             {\rm e}^{-b\left( r/R_{\rm e} \right)^{1/n}} \nonumber \\
  \rho_0 = {M\over L} \ I_0 \ b^{n(1-p)} \ {\Gamma(2n) \over
            2 R_{\rm e} \Gamma(n(3-p)) }, 
\end{array}
\end{equation}
where $r$ is the spatial radius and $\rho_0$ is the normalisation such
that the total mass from equation~(\ref{r2_app}, see Appendix A) equals that from
equation~(\ref{EqSer}).  We adopt Lima Neto et al.'s (1999) estimate,
or rather the updated value given in M\'arquez et al.\ (2000), for the
term $p$, for which a high--quality match between the exact,
deprojected S\'ersic profiles (solved numerically) and the above
expression is obtained when $p = 1.0 - 0.6097/n + 0.05563/n^2$, 
for $0.6<n<10$ and $10^{-2} \le R/R_{\rm e} \le 10^3$.  The
quantity ${M/L}\equiv \Upsilon$ is the mass--to--luminosity ratio,
which is typically taken not to depend on galaxy radius
but may of course do so.  The density at $r=R_{\rm e}$ is simply
$\rho_0{\rm e}^{-b}$.
Figure~(\ref{ldpf_PS}) shows the behaviour of this density profile for
different values of profile shape $n$.
Expressions and figures for the (enclosed) mass profile, the circular
velocity, and the spatial and line--of--sight velocity dispersion are
given in Appendix A for the spherical case.

Adding an inner power--law with slope $\gamma$, which is not to be
confused with the incomplete gamma function $\gamma(a,x)$, we obtain
the new model
\begin{equation} \label{cS2_1}
\begin{array}{ll}
  \rho(r)= \rho^{\prime} \left[1+\left({r_b\over r}\right)^{\alpha}\right]^
                       {\gamma / \alpha}
 \left\{  \left[ (r^{\alpha}+{r_b}^{\alpha})/ {R_{\rm e}}^{\alpha} \right]
       ^{-p/ \alpha} 
   {\rm e}^{-b \left[ (r^{\alpha} + r_b^{\alpha})/ R_{\rm e} ^{\alpha} \right]^{ 1/n\alpha }} 
 \right\}
 \nonumber \\
     \rho^{\prime} = \rho_b \ 2^{(p-\gamma) / \alpha} \ 
     \left(r_b\over R_{\rm e}\right)^p \
     {\rm e}^{ b\left( 2^{1 / \alpha} r_b / R_{\rm e} \right)^{1 / n} }.
\end{array}
\end{equation}
The break radius, $r_b$, denotes the transition where the profile
changes from one regime to the other, with $\rho _b$ the density at
this radius.  The parameter $\alpha$ controls the sharpness of the
transition.  For $r<<r_b$ equation~(\ref{cS2_1}) tends to a power--law
with slope $\gamma$.  For $r>>r_b$ equation~(\ref{cS2_1}) reduces to
equation~(\ref{r2_app}).

Modelling the light--profiles of luminous elliptical galaxies, 
Trujillo et al.\ (2004) have shown that the transition from the
inner ``core'' to the outer S\'ersic profile is probably rather sharp. 
Motivated by this result, we have chosen to make the 
transition between the inner power--law 
and the outer profile sharp by considering the $\alpha \to \infty$ limit.  
This reduces the above expression to a 5--parameter model 
capable of describing the entire radial extent of spherical stellar
systems with power--law ``cores'', and can be written as 
\begin{equation} \label{cS2_2}
\begin{array}{ll}
  \rho(r)= \rho_b \left[
\left({r_b\over r}\right)^{\gamma} h(r_b-r) +  {\bar \rho}
\left({r\over R_{\rm e}}\right)^{-p} {\rm e}^{ -b\left(r/R_{\rm e}\right)^{1/n} }
h(r-r_b)\right]
   \nonumber \\
{\bar \rho}=\left({r_b\over R_{\rm e}}\right)^{p} 
{\rm e}^{b\left(r_b / R_{\rm e}\right)^{1/n}}, 
\end{array}
\end{equation}
where $h(x)$ is the Heaviside step function such that
$h$=1 if $x>0$ and $h$=0 if $x\le0$.

\begin{figure*}
\begin{center}
\begin{minipage}{168mm}
\includegraphics[width=168mm]{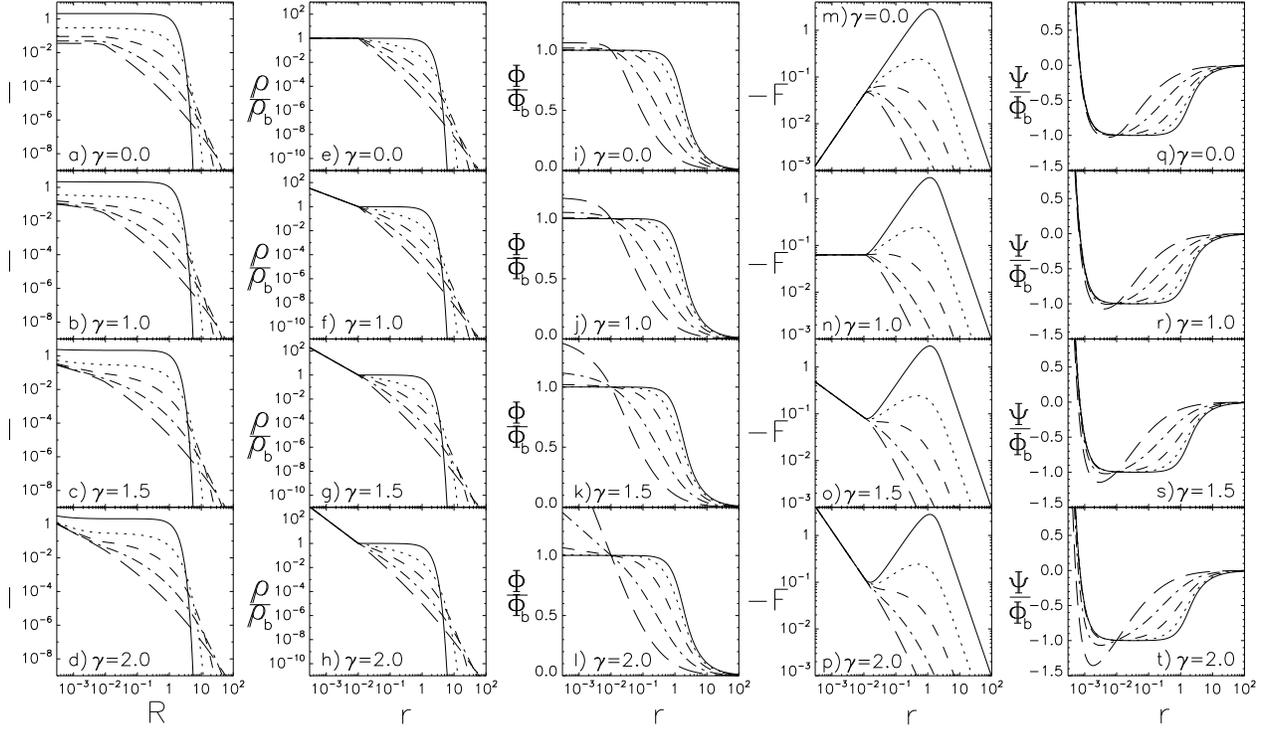}
\vskip30pt
\caption{Core galaxies. 
Panels a--d) show the projected intensity $I$ (equation~\ref{I_cS2}, exact 
numerical solution); 
panels i--l) show the potential (equation~\ref{pot_cS2} divided by $\Phi(r=r_b)$); 
panels m--l) show the force (equation~\ref{force_cS2}); and 
panels q--t) show the pseudo potential (equation~\ref{pseudo} 
divided by $|\Phi(r=r_b)|$ with $h^2=10^{-6}|\Phi(r=r_b)|$) 
associated with the density profiles (equation~\ref{cS2_2}) in panels e--h), 
for varying values of the outer profile shape 
$n$: $n=0.5$ (solid lines), $n=1$ (dotted), $n=2$ (dashed), 
$n=4$ (dash-dot), $n=10$ (double--dash); and varying central cusp slope $\gamma$. 
In these figures the scale radius $R_{\rm e}$ has a value of 1, as
does the scale density $\rho_b$ which occurs at a radius $r_b = 0.01R_{\rm e}$.
}
\label{ldpf_2}
\end{minipage}
\end{center}
\end{figure*}

These spherical models are illustrated in Fig.\ref{ldpf_2} e--h).  In each
panel a range of values of $n$ has been used, while the value of
$\gamma$ increases sequentially from 0 to 1, 1.5 and finally 2.
Depending on the parameter combination, one may have either a
partially depleted core or a central excess --- possibly representative
of a nuclear star cluster or adiabatic growth around a supermassive
black hole (e.g., Merritt 2004, and references therein). 
The break radius has been set to 0.01$R_{\rm e}$, which a) matches the
observed core--radii values of 0.01 to 0.02$R_{\rm e}$ found by
Trujillo et al.\ (2004, their table 2), and b) only uses the density
profile from Prugniel \& Simien (1997) over the radial range where it
provides a high--quality match to a deprojected S\'ersic profile.  For
reference, the nuclear star cluster ``half--width at half--maximum''
($HWHM$) values divided by their host bulge $R_{\rm e}$ values in
nucleated dwarf elliptical galaxies is 0.02 to 0.04 (Graham \&
Guzm\'an 2003, their table 2).

\subsection{Gravitational Potential and Force}

We derive an expression for the potential (assuming spherical symmetry) 
using the inverted form of the Poisson equation (e.g., 
Binney \& Tremaine 1987, their equation~2-22):
\begin{equation} \label{BT}
\Phi (r) = - 4 \pi G \left[ {1\over r} \int\limits_0^r \rho({\bar r}) {\bar r}^2
d{\bar r} + \int\limits_r^{\infty} \rho({\bar r}) {\bar r} d{\bar r} \right].
\end{equation}
After some algebraic manipulation (see Appendix B), one obtains
\begin{equation} \label{pot_cS2}
\Phi(r) =
{-4 \pi G}
\left\{
  \begin{array}{ll}
{1 \over r} J_1(r) + J_2(r) + L_1(r_b)
& \mbox{if $r \le r_b$}, \\
{1 \over r} J_1(r_b) + {1 \over r} L_2(r) + L_1(r)
& \mbox{if $r > r_b$},
  \end{array} \right.,
\end{equation}
where
\begin{equation} \label{J_1}
J_1(r) = \rho_b {r_b}^{\gamma}r^{3-\gamma}/(3-\gamma) \hskip50pt \mbox{if $\gamma < 3$}
\end{equation}
\begin{equation} \label{J_2}
J_2(r) = \rho_b {r_b}^{\gamma} \left\{
  \begin{array}{ll}
 {1 \over {2-\gamma}}
 \left({{r_b}^{2-\gamma}} - {r^{2-\gamma}} \right)
& \mbox{if $\gamma \ne 2$} \\
 \ln {r_b \over r} & \mbox{if $\gamma = 2$} 
\end{array} \right.,
\end{equation}
\begin{equation} \label{L_1}
L_1 (r) = \rho_b {\bar \rho} {R_{\rm e}}^2 n b^{n(p-2)}
  \Gamma\left(n(2-p),b\left({r \over R_{\rm e}}\right)^{1/n}\right),
\end{equation}
\begin{equation} \label{L_2}
L_2 (r) = \rho_b {\bar \rho} {R_{\rm e}}^3 n b^{n(p-3)} \left[
   \Gamma \left(n(3-p),b\left({r_b \over R_{\rm e}}\right)^{1/n}\right)
  -\Gamma \left(n(3-p),b\left({r \over R_{\rm e}}\right)^{1/n}\right)
  \right], 
\end{equation}
and 
\begin{equation}
\Gamma(a,x) = \int_x^{\infty}{\rm e}^{-t}t^{a-1} dt, \hskip10pt  a>0 
\end{equation}
is the complement to the incomplete gamma function shown in 
equation~(\ref{eqgam}).  
Although $J_1$ diverges when $\gamma=3$, this is unlikely to be 
a problem because such steep inner profile 
slopes are not observed in real galaxies (e.g. Gebhardt et al.\ 1996; 
Milosavljevi\'c et al.\ 2002; Ravindranath, Ho, \& Filippenko 2002). 
Fig.\ref{ldpf_2}i--\ref{ldpf_2}l show the gravitational potential of 
this model as a function of spatial radius $r$.  
The case when the break radius equals zero is shown in Fig.~(\ref{ldpf_PS}c). 

The central potential $\Phi (0) = -4\pi G[J_2(0) + L_1(r_b)]$, 
for $\gamma \ne 2$.  When $r_b=0$, $\Phi (0) = -4\pi G L_1(0)$, 
with $\rho_b {\bar \rho} = \rho_0$.

The corresponding radial force is computed by differentiation with respect to 
the spatial radial coordinate $r$.  After some algebraic cancellations
(see Appendix B), the force is found to be
\begin{equation} \label{force_cS2}
F(r) = - {{d \Phi} \over dr} = {4 \pi G}
\left\{
  \begin{array}{ll}
-{1 \over r^2} J_1(r) 
& \mbox{if $r \le r_b$}, \\
-{1 \over r^2} J_1(r_b) - {1 \over r^2} L_2(r)
& \mbox{if $r > r_b$},
  \end{array} \right..
\end{equation}
We remind readers of the logarithmic radial scale used in
Fig.~\ref{ldpf_2}, and caution that the gradient to the curves shown in
Fig.\ref{ldpf_2}i--\ref{ldpf_2}l should be interpreted with care when
comparing them with the force shown in Fig.\ref{ldpf_2}m--\ref{ldpf_2}p.
The case when the break radius equals zero is shown in Fig.~(\ref{ldpf_PS}d).

In order to carry out orbital integration, one needs to compute the
force along an orbit.  For a fixed break radius $r_b$, the term in the
first half of the expression $L_2(r)$ is constant and therefore only
needs to be computed once at the beginning of a simulation.
Therefore, the incomplete gamma function $\Gamma(a,x)$ need only be
called once per evaluation of the force (the term in the second half
of $L_2(r)$), and not at all if the star is inside the break radius,
i.e.\ $r < r_b$.

The abrupt change in the force --- whose nature is not only dependent on
the value of the inner profile slope $\gamma$ and the S\'ersic index $n$ 
but also on the ratio $r_b/R_{\rm e}$ --- is not a
problem for constructing models because the force remains both
negative and continuous at $r_b$.
Orbit stability can be checked using the ``pseudo'' potential $\Psi(r)$, 
sometimes also referred to as the ``effective'' potential 
(Landau \& Lifshitz 1976; Goldstein 1980), and given by 
\begin{equation}
\Psi (r) = \Phi (r) + \frac{h^2}{2r^2}, \label{pseudo}
\end{equation}
where $h$ is the angular momentum per unit mass. 
Uisng $h^2=10^{-6}|\Phi(0.01R_{\rm e})|$, and normalising $\Psi(r)$ by
dividing by $|\Phi(0.01R_{\rm e})|$, in order to highlight orbits near
the radius $0.01R_{\rm e}$, Fig.\ref{ldpf_2}q--\ref{ldpf_2}t reveals 
that the abrupt change in density and force at $r_b$ does not
result in a double minimum for the pseudo potential --- which would be
indicative of an unstable configuration.  Instead, for a given orbital
energy less than zero, particles remain bound within the curve defined
by the pseudo potential.

In summary, we have a set of equations consisting of analytical
expressions, and one special function for which fast computer codes
exist, that can a) simulate the range of observed light--profile shapes
and b) model partially depleted cores and/or certain additional nuclear 
components.

\subsection{Projected Surface Brightness Profile \label{secSB}}

The projection of the above density profile, i.e.\ the
intensity--profile or ``light--profile'', is computed by solving the
Abel integral
\begin{equation} \label{rho2I}
I(R)={2\over {\Upsilon}} \int\limits^{\infty}_R
 {{\rho(r) r}\over{\sqrt{r^2-R^2}}} dr, 
\end{equation}
where $R$ is the projected radius and $\Upsilon$ is the
mass--to--luminosity ratio.  If equation~(\ref{cS2_2}) is rewritten as
$\rho(r)=\rho_1(r) + \rho_2(r)$, where
\begin{eqnarray} \label{rhos}
\rho_1(r) &=& \rho_b \left({r_b\over r}\right)^{\gamma} h(r_b-r), 
              \hskip10pt {\rm and} \nonumber \\
\rho_2(r) &=& \rho_b {\bar \rho}\left({r\over R_{\rm e}}\right)^{-p} 
{\rm e}^{-b\left(r/R_{\rm e}\right)^{1/n}} h(r-r_b),
\end{eqnarray}
then the corresponding intensity can be expressed as 
\begin{equation} \label{I_cS2}
  I(R) = \left\{
  \begin{array}{ll}
{2\over \Upsilon} \int\limits^{r_b}_R {{\rho_1(r) r}\over{\sqrt{r^2-R^2}}} dr +
{2\over \Upsilon} \int\limits^{\infty}_{r_b} 
{{\rho_2(r) r}\over{\sqrt{r^2-R^2}}} dr \equiv I_1(R) + I_2(R;r_b) & 
\mbox{if $R \le r_b$,} \\
{2\over \Upsilon} \int\limits^{\infty}_{R} 
{{\rho_2(r) r}\over{\sqrt{r^2-R^2}}} dr \equiv I_2(R) & \mbox{if $R > r_b$.}
  \end{array} \right.
\end{equation}

As shown in Appendix B, the first term $I_1(R)$ is such that 
\begin{equation} \label{I1_cS2}
  I_1(R) = {2 \over \Upsilon} \rho_b {r_b}^{\gamma} \left\{
  \begin{array}{ll}
\sqrt{r_b^2 -R^2} & \mbox{if $\gamma=0$,}\\
\ln{{r_b+\sqrt{r_b^2 -R^2}}\over R} & \mbox{if $\gamma=1$,}\\
R^{-1} \sin^{-1}\sqrt{1-R^2/r_b^2} & \mbox{if $\gamma=2$,}\\
{1\over 2} R^{1-\gamma} B_{1-R^2/{r_b}^2} \left({1 \over 2},
{{\gamma-1}\over{2}}\right)
& \mbox{otherwise,}
\end{array} \right.
\end{equation}
with $B$ the incomplete Beta function defined as
\begin{equation} \label{BETA}
B_{x} (y,z) = \int\limits_0^x u^{y-1} (1-u)^{z-1} du.
\end{equation}

\begin{figure}
\begin{center}
\includegraphics[width=8cm]{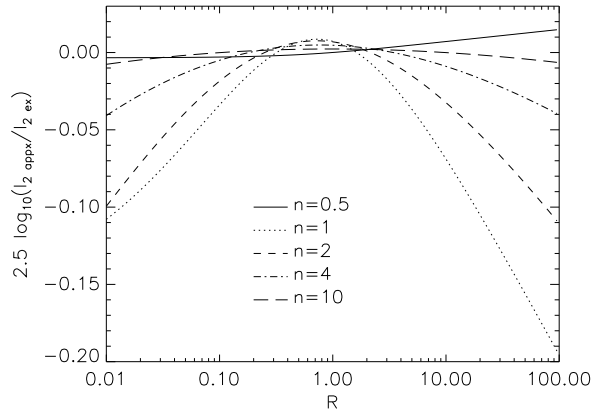}
\caption{
The relative error between the term $I_2(R)$ given in 
equation~(\ref{I_cS2}) and the approximation given in equation~(\ref{mm}) 
is shown as a function of radius, normalised by the effective radius $R_{\rm e}$, 
for varying light--profile shapes $n$. 
}
\label{error_I2}
\end{center}
\end{figure}

The only expressions we do not provide an exact equation for 
are $I_2(R)$ and $I_2(R; r_b)$.  However, because 
Prugniel \& Simien (1997) devised equation~(\ref{r2_app}) 
to match the deprojected form of the S\'ersic light--profile, 
it makes sense to use the $R^{1/n}$ model as a suitable approximation. 
One therefore has that 
\begin{equation} \label{mm}
I_2(R) \approx I_{\rm e} {\rm e}^b {\rm e}^{-b\left(R/R_{\rm e}\right)^{1/n}}
\end{equation} 
with $I_{\rm e}$ the (projected) intensity at the (projected) radius
$R_{\rm e}$.  The value of $I_{\rm e}$ can be set in terms of the
density model parameters if one applies the condition that the total
mass given by equation~(\ref{mm}, see equation~\ref{Sertot}) equals
the total mass from equation~(\ref{r2_app}, see Appendix A), yielding
\begin{equation} 
I_{\rm e} = { {2 {\rm e}^{-b} R_{\rm e} \rho_b {\bar \rho} \Gamma(n(3-p))} \over
{\Upsilon \Gamma(2n) b^{n(1-p)}}}, 
\end{equation} 
with ${\bar \rho}$ given in equation~(\ref{cS2_2}). 
Over the radial interval $10^{-2} \le R/R_{\rm e} \le 10^2$, the
maximum difference in surface brightness between this approximation
for $I_2(R)$ and the exact value is about 0.1 mag arcsec$^{-2}$ if
$n>2$, and only 0.04 mag arcsec$^{-2}$ when $n=4$
(Fig.~\ref{error_I2}).  For a Gaussian profile ($n=0.5$) the agreement
is very good, while for an exponential profile ($n=1$) the match at
large radii is rather poor --- increasing to $\sim$0.2 mag
arcsec$^{-2}$ at $R/R_{\rm e}=100$.
%
%

The only term requiring numerical evaluation is $I_2(R;r_b)$,
although, if desired, one may choose to additionally evaluate $I_2(R)$
numerically rather than using the analytical approximation given in
equation~(\ref{mm}). 
Given that the intensity profile is likely only computed at the end, 
or at a few intermediate stages, of a simulation, the computation 
time involved in deriving the light--profile is not a concern. 

The exact (numerically computed) projected light--profiles associated
with the density models in Fig.~(\ref{ldpf_PS}b) are shown in 
Fig.~(\ref{ldpf_PS}a), and those in 
Fig.\ref{ldpf_2}e--\ref{ldpf_2}h are
shown in Fig.\ref{ldpf_2}a--\ref{ldpf_2}d.
In passing we note that when $\gamma=0$ and $R<r_b$, the
slope of the surface brightness profile is close to but not exactly
zero (Fig.\ref{ldpf_2}a).
We also note that the value of $I_{\rm e}$ and $R_{\rm e}$ are stable
against increases in $r_b$ and changes in $\gamma$ because they
pertain to the outer, undisturbed profile when there is a partially
depleted core, and to the underlying host galaxy profile when a
central flux excess is present.  As such, when $r_b>0$, $I_{\rm e}$ and
$R_{\rm e}$ are not the total galaxy half--light values.  In practice,
the stellar flux deficit is only $\sim$0.1\% for ``core'' galaxies
(Graham 2004) and the stellar flux excess $\sim$1\% in nucleated dwarf
elliptical galaxies (Graham \& Guzm\'an 2003), and so the actual
discrepancy is not greater than $\sim$1\%.

\section{Application to real galaxy profiles}

Over a quarter of a century ago, de Vaucouleurs \& Capaccioli (1979)
demonstrated that, when the inner 10 arcseconds were excluded, 
the $R^{1/4}$ model provided a very good fit to the
light--profile of NGC~3379 ($M_B \sim -20$ mag). 
Since then, Capaccioli and his collaborators have shown that different
galaxies can be equally well fit, but only by using models with a very
different curvature (e.g., Caon et al.\ 1993).
Bertin, Ciotti \& Del Principe (2002) have shown $R^{1/n}$ fits to
the circularised light--profiles of four (illustrative) galaxies from
Caon et al.\ (1993).  Here we deproject these galaxies' major--axis
light--profiles and fit the Jaffe, Hernquist and Dehnen models, and
explore how well they describe the luminosity--density profile in
comparison with the model of Prugniel \& Simien (1997) given in 
equation~(\ref{r2_app}).
In addition, we include the \textit{HST}--resolved dwarf elliptical
galaxy\footnote{LGC: Leo Group Catalog.} LGC~47 (Stiavelli et al.\
2001), and the ``core'' galaxy NGC~3348 (Rest et al.\ 2001; Trujillo
et al.\ 2004) to which we apply our new core-density model
(equation~\ref{cS2_2}).

\subsection{S\'ersic and core--S\'ersic model}

\begin{figure*}
\begin{center}
\begin{minipage}{160mm}
\includegraphics[width=6.8cm,angle=270]{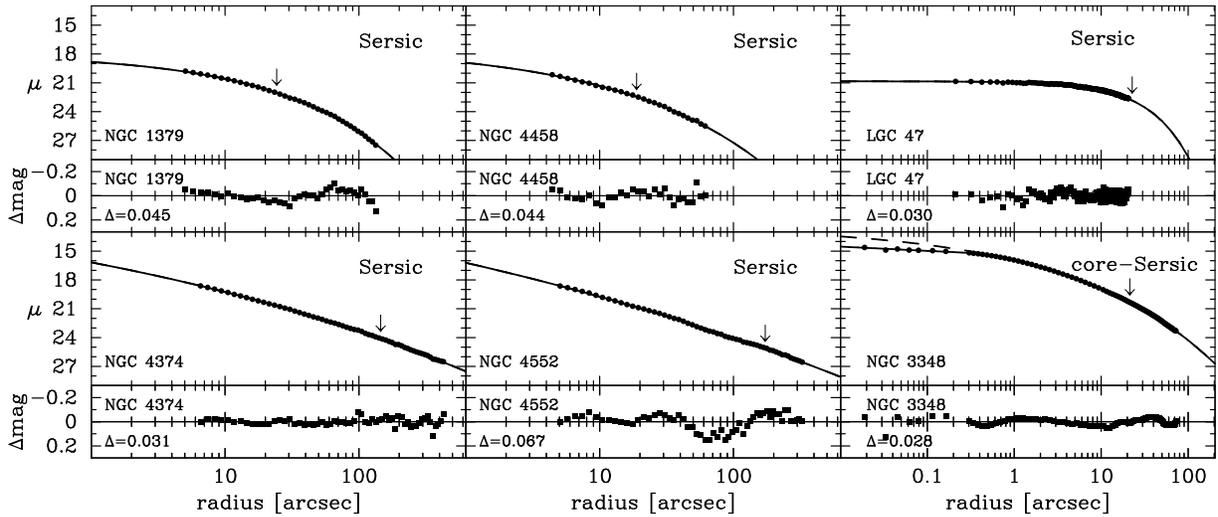}
\caption{
Best--fitting S\'ersic model to the major--axis, $B$--band
light--profiles of NGC~1379, 4458, 4374 and 4552
(data from Caon et al.\ 1990; Caon et al.\ 1993), 
and the major--axis, $I$--band light--profile of LGC~47 (data from
Stiavelli et al.\ 2001). 
The best--fitting core--S\'ersic model to the major--axis, $R$--band
light--profile of the ``core'' galaxy NGC~3348 is also shown 
(data from Trujillo et al.\ 2004); the inward extrapolation of the 
outer S\'ersic profile is shown by the dashed curve. 
The model parameters are given in Table~\ref{Tab_Ser}. 
The effective half--light radii are marked with an arrow. 
}
\label{bomba}
\end{minipage}
\end{center}
\end{figure*}

Fig.\ref{bomba} shows the S\'ersic (1963, 1968) model applied to
the ground--based,  major--axis, $B$--band 
light--profiles of NGC~1379, 4458, 4374, and 4552, and applied to the 
\textit{HST}--based,  major--axis, $I$--band light--profile of LGC~47. 
The data have come from Caon et al.\ (1993) and 
Stiavelli et al.\ (2001), respectively.
Also shown is the core--S\'ersic model (Graham et al.\
2003b; Trujillo et al.\ 2004) applied to the \textit{HST}--based, major--axis, 
$R$--band light--profile of NGC~3348.  Table~\ref{Tab_Ser}
shows the best--fitting parameters.  For the first four galaxies, the
values agree with those reported in Caon et al.\ (1993), with the
exception that we find a slightly smaller value of $n$ for NGC~4552.
However, the difference between an $n=12$ and $n=14$ profile is minimal.  
For LGC~47, Stiavelli et al.\ (2001) presented fits to the 
geometric mean ($r=\sqrt{ab}$), $V$--band light--profile, finding $n\sim 1.5$. 
Due to possible colour gradients, and ellipticity gradients, it is 
expected that our fit to the major--axis, $I$--band light--profile may be
slightly different: nonetheless, we derived a similarly small value 
of $n=1.1$. 
NGC~1379 and NGC~4458 have values of $n=2.0$ and 2.6 respectively.
NGC~4374 and NGC~4552 have values of $\sim 8$ and $\sim 12$,
considerably greater than 4.  On the other hand, NGC~3348 has a value
of $n$ close to 4, but possesses a distinct core.  Although it is
likely the large galaxies NGC~4374 and NGC~4552 may also possess a
partially depleted core, the inner few arcseconds have been excluded
from their profile due the effects of seeing.

\begin{table}
\caption{S\'ersic and core--S\'ersic parameters from the fits 
in Fig.\ref{bomba}.\label{Tab_Ser}} 
\begin{tabular}{@{}lcccccccc@{}}
Gal  & Dist. & Band & $\mu_{\rm e}$ & $R_{\rm e}$ & $n$  & $\gamma$ & $R_b$ & $\mu_b$\\
Id.\ & mod. &  & mag arcsec$^{-2}$ &  $^{\prime\prime}$  & & & $^{\prime\prime}$ & mag arcsec$^{-2}$ \\
\hline
NGC 1379 & 31.51 & $B$ &  22.02     &    24.3     &  2.0 &   ...    &  ...  &  ...   \\
NGC 4458 & 31.18 & $B$ &  22.46     &    18.9     &  2.6 &   ...    &  ...  &  ...   \\
LGC 47   & 30.30 & $I$ &  22.84     &    22.7     &  1.1 &   ...    &  ...  &  ...   \\
NGC 4374 & 31.32 & $B$ &  24.11     &    146      &  8.2 &   ...    &  ...  &  ...   \\
NGC 4552 & 30.93 & $B$ &  25.16     &    172      & 11.8 &   ...    &  ...  &  ...   \\
NGC 3348 & 33.08 & $R$ &  ...       &    21.4     &  3.8 &   0.18   &  0.43 &  15.30 \\
NGC 2986 & 32.31 & $R$ &  ...       &    75.7     &  6.7 &   0.25   &  0.77 &  15.60 \\
NGC 4291 & 32.09 & $R$ &  ...       &    17.8     &  5.3 &   0.14   &  0.38 &  14.54 \\

\hline
\end{tabular}
\end{table}

\subsection{Density profiles} 

In order to obtain each galaxy's deprojected light--profile, a
non--parametric deprojection was used when solving the appropriate
Abel integral.  The best--fitting light--profile models were however
used to extrapolate the observed light--profiles to infinity.  
This simply meant using the best-fitting S\'ersic model to 
extend the observed data to larger radii in order for one to
compute the deprojected light--profile, i.e.\ the density profile. 
The actual choice of extrapolation only affects the outer 
luminosity--density profile slightly. 
%
Due to the coarser radial sampling of the ground--based profiles than
the \textit{HST} profiles, their deprojected profiles are somewhat
noisier.  We applied box--cart smoothing with four different box sizes
and refitted the various density models; in every instance the
best--fitting model parameters where practically identical, suggesting
such noise is not an issue.

In the case of the dwarf elliptical galaxy LGC~47, its observed
(exponential) light--profile only went out to 1 $R_{\rm e}$, not
enough to properly show how such stellar distributions decline at
large radii.  For this galaxy only, we model the density profile to
radii larger than observed.  Given that galaxies, and bulges, with
exponential profiles do exist, we felt that the analysis of such a
profile would be of value. 

The luminosity--density profiles (Fig.\ref{Jaffe}--\ref{new_mod}) have
been calibrated in units of $L_{\sun}$ pc$^{-3}$ using the distance
moduli provided in Tonry et al.\ (2001) for the Caon et al.\ sample,
and a distance of 41 Mpc for NGC~3348 and 11.5 Mpc for LGC~47, taken
from their respective papers.  We used solar absolute magnitudes of 
$M_B$=5.47, $M_R$=4.28 and $M_I$=3.94 (Cox 2000). 
A Hubble constant $H_{\rm 0}=75$ km s$^{-1}$ Mpc$^{-1}$ was used. 

The value of $\Delta$ associated with the residual profiles shown in
Fig.\ref{Jaffe}--\ref{new_mod} is not quite the rms error, but is
computed using the expression
\begin{equation}
\Delta = \frac{\sqrt{\sum_{i=1}^m {\delta_i}^2}}{m-k}, 
\end{equation}
where $m$ is the number of data points, $\delta_i$ is the $i$ th
residual and $k$ is the number of parameters in the fitted model.
Because the different models have different numbers of free
parameters, this provides a more appropriate measure for model
comparison.

\subsection{Jaffe model}

\begin{figure*}
\begin{center}
\begin{minipage}{160mm}
\includegraphics[width=6.8cm,angle=270]{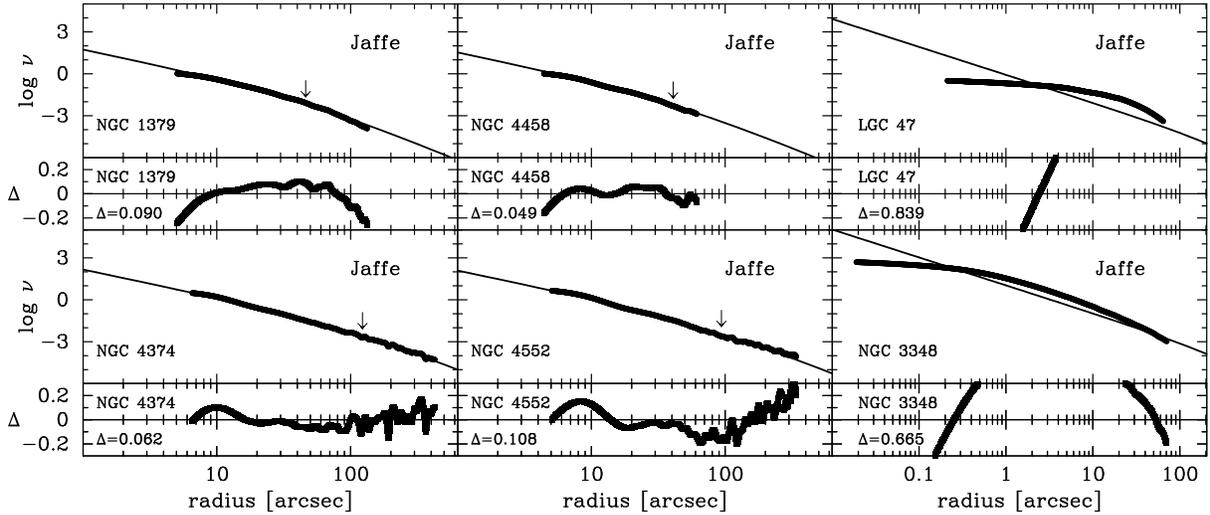}
\caption{
Best--fitting Jaffe model to the deprojected light--profiles
of the galaxies in Fig.\ref{bomba}.  The luminosity--density 
profile $\nu(r) = \rho(r) / (M/L)$ 
is in units of $L_{\sun,B}$ pc$^{-3}$ for NGC~1379, 4458, 4374
and 4552, $L_{\sun,R}$ pc$^{-3}$ for NGC~3348, 
and $L_{\sun,I}$ pc$^{-3}$ for LGC~47.
The model parameters are given in Table~\ref{Tab_Mod}. 
The arrow denotes the size of the scale radius $a$. 
}
\label{Jaffe}
\end{minipage}
\end{center}
\end{figure*}

Fig.\ref{Jaffe} presents the best--fitting (2--parameter) Jaffe (1980)
model, $\rho (r) = [4 \rho(a)](r/a)^{-2}(1+r/a)^{-2}$, to each
galaxy's luminosity--density profile.  The scale--length is denoted by
$a$, and $\rho(a)$ is simply the density at $r=a$.  The position of
the transition radius, i.e.\ the scale--length, where this model changes
from an inner power--law slope of $-$2 to an outer power--law slope of
$-$4 is marked with an arrow in Fig.\ref{Jaffe}.  In the case of
LGC~47 and NGC~3348, the Jaffe model is clearly inadequate to describe
the stellar profile.  We also note that the best--fitting models have
transition radii well beyond the observed radial extent of the
galaxies.  The hump--shaped residual profiles (upper panels) for the
low $n$ galaxies NGC~1379 and NGC~4458, and the bowl--shaped residual
profiles for the high $n$ galaxies NGC~4374 and NGC~4552 are clear
indications that the Jaffe model is unable to match the global
curvature in these galaxies' stellar distributions.

\subsection{Hernquist model} 

\begin{figure*}
\begin{center}
\begin{minipage}{160mm}
\includegraphics[width=6.8cm,angle=270]{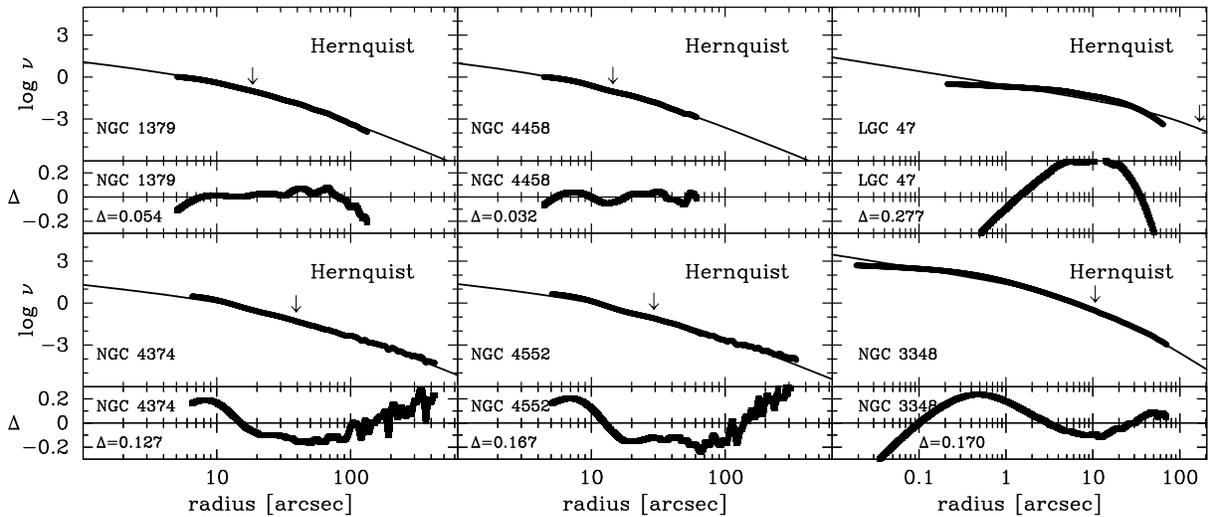}
\caption{
Same as Fig.\ref{Jaffe}, except that the best--fitting
Hernquist models are shown. 
}
\label{Hernquist}
\end{minipage}
\end{center}
\end{figure*}

In Fig.\ref{Hernquist} one can see the best--fitting (2--parameter)
Hernquist (1990) models, $\rho (r) = [8
\rho(a)](r/a)^{-1}(1+r/a)^{-3}$.  This model is also incapable of
describing the core galaxy NGC~3348, and the hump/bowl shaped residual
profiles of the other galaxies largely reflects the failure seen with
the Jaffe model.  We do however note that in the case of NGC~4458, the
(non-core) galaxy with the closest S\'ersic index to a value of 4, the
fit is quite acceptable. 

\subsection{Dehnen model} 

The (3--parameter) Dehnen (1993) model, $\rho(r) = [2^{4-\gamma} \rho(a)]
(r/a)^{-\gamma}(1+r/a)^{\gamma-4}$, which subsumes the first two
models as special cases, does noticeably better.  This is due to the
fact that it has an additional (third) parameter over the previous two
models that actually measures the inner power--law slope, rather than
assuming some fixed value.  Obviously the Dehnen model fails to fit
the detailed profile of NGC~3348.  In a sense, it also fails to match
NGC~4552, setting the transition radius to 500$^{\prime\prime}$ --- an
arbitrary large upper limit in our code --- well outside of the
observed radial range.  The reason for this large transition radius is
however understood.  Graham \& Driver (2005) show that for large
values of $n$, the S\'ersic model tends to a power--law such that
$I(R)\sim R^{-2}$; which results in $\rho(r)\sim r^{-3}$.  NGC~4552
has the largest value of $n$ in our sample, and the optimal Dehnen
model is simply a (single) power--law with slope $\gamma$.  

\begin{figure*}
\begin{center}
\begin{minipage}{160mm}
\includegraphics[width=6.8cm,angle=270]{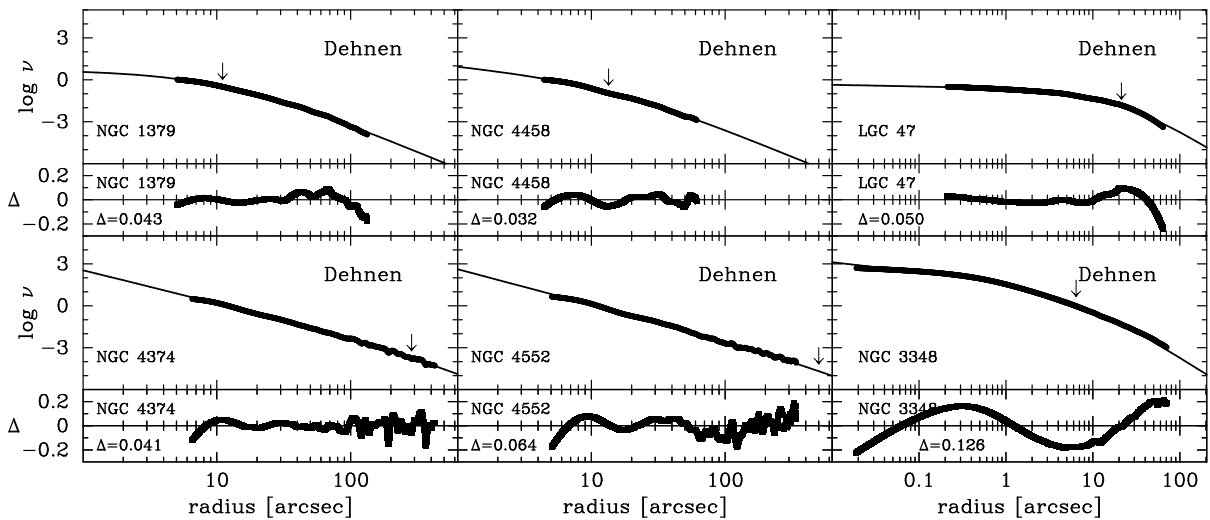}
\caption{
Same as Fig.\ref{Jaffe}, except that the best--fitting
Dehnen models are shown. 
}
\label{Dehnen}
\end{minipage}
\end{center}
\end{figure*}

In the case of the lower $n$ galaxies (LGC~47 and NGC~1379), one can
see that the Dehnen model --- constructed to match $R^{1/4}$
light--profiles --- does not decline as quickly as the observed
structures.  Due to the fact that we used an exponential--like model ($n$=1.1)
to extrapolate the light--profile of LGC~47, this eliminates the
possibility that the observed mismatch at large radii may be unique to
this galaxy.  Rather, the mismatch reflects the well known fact that
exponential light--profiles have outer density profiles that decline
more quickly with radius than $r^{-4}$.  Not surprisingly, an almost
identical result to that seen in Fig.~\ref{Dehnen} for LGC~47 is
obtained when we deproject a pure $n=1$ exponential light--profile.
As a rule, the lower the value of the S\'ersic index $n$, the steeper
the outer decline in density and the greater the mismatch with the 
Dehnen model.

\begin{table*}
\begin{minipage}{170mm}
\caption{Best--fitting parameters from a range of density models.
The units are $L_{\sun}$ pc$^{-3}$ for $\rho$ and arcseconds for $a$. 
}
\label{Tab_Mod}
\begin{tabular}{lcccccccccccccc}
\hline
Gal. & Band & \multicolumn{2}{c}{Jaffe} & \multicolumn{2}{c}{Hernquist} & \multicolumn{3}{c}{|----- Dehnen -----|} &
\multicolumn{6}{c}{|-------------------- Equation~(\ref{cS2_2}) --------------------|} \\
Id.\  &  &  a  & $\log \rho(a)$   &  a  & $\log \rho(a)$  &  a  & $\log \rho(a)$ & $\gamma$  &  
$R_{\rm e}$  & $\log \rho(R_{\rm e})$  &  $n$  & $\gamma$ & $r_b$ & $\log \rho(r_b)$ \\
\hline
NGC 1379 & $B$ & 46.2 & -2.17 & 18.5 & -1.03 & 11.1 & -0.49 & 0.00 & 24.7 & -1.31 &  2.1 & ...  & ...  &  ... \\
NGC 4458 & $B$ & 40.8 & -2.27 & 14.5 & -0.99 & 13.5 & -0.91 & 0.89 & 18.8 & -1.28 &  2.5 & ...  & ...  &  ... \\
LGC 47   & $I$ & 500  & -6.07 & 168  & -3.71 & 21.3 & -1.91 & 0.12 & 22.4 & -1.85 &  0.9 & ...  & ...  &  ... \\
NGC 4374 & $B$ & 123  & -2.61 & 39.2 & -1.16 & 285  & -3.74 & 2.36 & 131  & -2.73 &  7.7 & ...  & ...  &  ... \\
NGC 4552 & $B$ & 93.9 & -2.45 & 29.3 & -0.97 & 500  & -4.68 & 2.54 & 168  & -3.21 & 10.8 & ...  & ...  &  ... \\
NGC 3348 & $R$ & 500  & -4.97 & 10.6 &  0.46 & 06.4 &  0.14 & 0.71 & 20.2 & ...   &  3.6 & 0.44 & 0.37 & 2.15 \\ 
NGC 2986 & $R$ & ...  & ...   & ...  & ...   & ...  & ...   & ...  & 79.1 & ...   &  6.5 & 0.71 & 0.78 & 1.92 \\
NGC 4291 & $R$ & ...  & ...   & ...  & ...   & ...  & ...   & ...  & 21.4 & ...   &  5.6 & 0.43 & 0.46 & 2.56 \\
\hline
\end{tabular}
\end{minipage}
\end{table*}

\subsection{Prugniel--Simien density profile, 
and our adaptation for galaxies with power--law cores}

\begin{figure*}
\begin{center}
\begin{minipage}{160mm}
\includegraphics[width=6.8cm,angle=270]{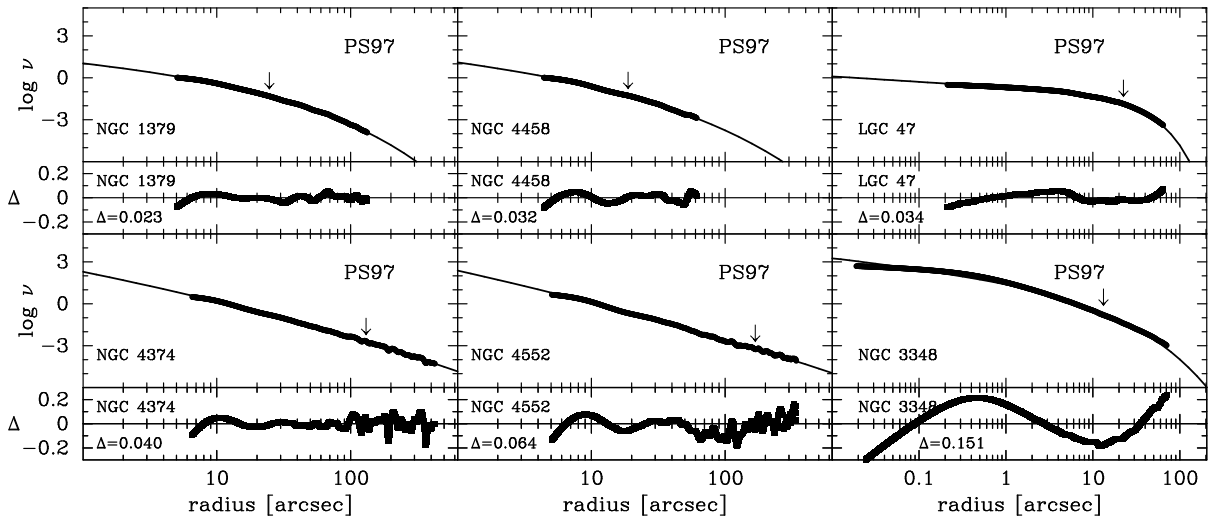}
\caption{
Same as Fig.\ref{Jaffe}, except that the best--fitting
model of Prugniel \& Simien (1997), as given by equation~(\ref{r2_app}), 
is shown here. 
The arrow marks the radius $R_{\rm e}$.  
%
}
\label{PS_mod}
\end{minipage}
\end{center}
\end{figure*}

Fig.\ref{PS_mod} presents the best--fitting (3--parameter) model
of Prugniel \& Simien (1997). 
Aside from NGC~3348, 
the fits are obviously rather good, with only NGC~4552 suggesting the
presence of additional fine structure that has been missed.  
In fact, Caon et al.\ (1993) classified this as a lenticular galaxy; 
the other galaxies are elliptical, i.e. they do not have an embedded 
large--scale disk. 
With the exception of the core galaxy NGC~3348, 
every fit is equal to or better than those obtained with the (3--parameter) 
Dehnen model.  This is especially the case for the low $n$ galaxies 
LGC~47 and NGC~1379.  

\begin{figure*}
\begin{center}
\begin{minipage}{160mm}
\includegraphics[width=3.8cm,angle=270]{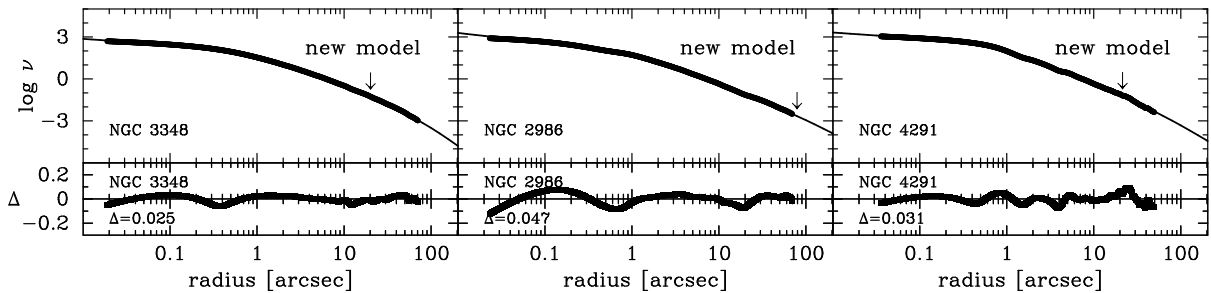}
\caption{
New density model (equation~\ref{cS2_2}) applied to 
the core galaxies NGC~3348, 2986 and 4291.
The arrow marks the model radius $R_{\rm e}$.  
%
}
\label{new_mod}
\end{minipage}
\end{center}
\end{figure*}

Fig.\ref{new_mod} presents the new
(5--parameter) model (equation~\ref{cS2_2}) applied to the core galaxy
NGC~3348.  The model performs well and is the only density model we
are aware of that can fit galaxies possessing partially depleted cores.
The best--fitting parameters are given in Table~\ref{Tab_Mod}.
Although the luminosity--density at a spatial radius $r$ equal to
$R_{\rm e}$ is not a formal parameter of the new model, it is equal to
$10^{-1.26} L_{\sun,R}$ pc$^{-3}$ for NGC~3348.

We have also included two additional ``core'' galaxies: NGC~2986
and NGC~4291.  The best--fitting core-S\'ersic parameters derived here
for NGC~4291 are the same as those given in Trujillo et al.\ (2004),
as was the case with NGC~3348.  However, our core--S\'ersic parameters
for NGC~2986 are somewhat different; the reason is due to the fact
that we exclude the (highly deviant) outermost two data points from the
light--profile given in Trujillo et al., which may be due to poor
sky--subtraction.  Although the core--parameters did not change much,
the value of $n$ increased by 25\% and $R_e$ increased from 44 to 76 
arcseconds.

NGC~2986 and NGC~4291 were not shown in the previous figures because
they had residual profiles very similar to that seen for NGC~3348.
They are included here so one can see the new model's applicability to
other core galaxies.  Interestingly, the residual profiles seem to
suggest that the core density profiles have only approximately a
power--law structure.

\section{Discussion}

Just like the $R^{1/4}$ model, the $R^{1/n}$ model was offered as an
empirical fitting function.  A physical basis for its origin is not
widely recognised.  M\'arquez et al.\ (2000, 2001, and references
therein) have argued it is a natural result from the (near)
conservation of specific entropy.  Other theoretically motivated
models, based on the statistical mechanics of partially complete
violent relaxation, also lead to density profiles that match
deprojected S\'ersic $R^{1/n}$ profiles, at least for $2.5 < n <
8.5$ (Trenti \& Bertin 2005, see also Hjorth \& Madsen 1995).  We do
not, however, attempt to solve this question here, but do note that
the Jaffe, Hernquist and Dehnen models, like the new model presented
here, are simply useful empirical fitting functions.

Fig.\ref{Jaffe} and \ref{Hernquist} clearly reveal that the Jaffe and
Hernquist models fail to describe the stellar distribution of low-- and
high--$n$ galaxies, relative to $n=4$.  This is not particularly 
surprising as they were only designed to match galaxies with an $n=4$
light--profile.  
The Dehnen model does better, but fails to model the lower $n$
galaxies in our sample.  This can be seen in the outer part of the
profile for NGC~1379 and LGC~47 in Fig.~\ref{Dehnen}, where the
density declines more rapidly with radius than $r^{-4}$.  At the
high--$n$ end, the (projected) intensity profile $I(r) \rightarrow
r^{-2}$ as the S\'ersic index $n \rightarrow \infty$ (Graham \& Driver
2005), and the density profile tends towards a power-law with
$\rho(r) \rightarrow r^{-3}$.  For reference, the isothermal model has
$\rho(r) \propto r^{-2}$.  Ignoring the presence of partially 
depleted cores, for galaxies with values of
$n$ greater than $\sim$8, a single power--law can approximate
their density profiles and the Dehnen model appears to work well,
albeit by setting the scale radius, $a$, to very large radii.


Zhao (1996) developed a 5--parameter generalisation of the Dehnen
model such that both the inner and outer power--law slope can be
adjusted, along with the radius, density and sharpness of the
transition region.  The model was first introduced by Hernquist (1990,
his equation 43) and has the same structural form as the ``Nuker''
model (e.g., Lauer et al.\ 1995).  However, as explained in Graham et
al.\ (2002, 2003b), such double power--law models are not appropriate
for describing profiles with obvious logarithmic curvature, that is,
profiles without inner and outer power--laws but whose slopes
continuously vary as a function of radius --- which is the case for
the luminosity--density profiles of most elliptical galaxies and
bulges.  Although in the absence of a partially depleted stellar core
this model should be able to approximate real luminosity--density
profiles, the parameters themselves are highly sensitive to the fitted
radial range.
This is because the parameters
simply adjust themselves in order to match the curvature in whatever
part of the profile one includes in the fit, providing inner and 
outer power--law slopes that are a product of the radial range sampled. 
Consequently, this model is not explored here. 
The 3--parameter model of Prugniel \& Simien (1997) was, however,
designed to describe profiles with curvature and does therefore not suffer 
from such a problem.

The density profile of Prugniel \& Simien (equation~\ref{r2_app}) 
may prove helpful for 
describing gravitational lenses (e.g., Cardone 2004; Kawano et al.\
2004) and the dark matter distributions of halos built in $\Lambda$CDM
cosmological simulations.  Demarco et al.\ (2003) have already
observed the hot gas in galaxy clusters to have a (projected) S\'ersic
distribution.  Intriguingly, the curved nature of simulated dark
matter profiles has resulted in the recent discovery that S\'ersic's
model describes them better than the generalised NFW\footnote{If
cosmological simulations of dark matter halos indeed have density
profiles which are logarithmically curved, i.e.\ if they have slopes
which vary continuously with radius (e.g. Navarro et al.\ 2004; Hansen
\& Moore 2005), then the issue of whether the central cusp slope is -1
or -1.5 or smaller may need to be reconsidered.} 
double power--law model with inner and outer slopes of $-$1
and $-$3, respectively (Merritt et al.\ 2005).  Whether or not they
were aware of the fact, Navarro et al.\ (2004; their equation 5) and
Cardone, Piedipalumbo \& Tortora (2005; their equation 10) presented
S\'ersic's model as a leading candidate to describe the density
profiles of dark matter halos.  It would be of interest to investigate
whether the generalisation of the Mellier--Mathez (1987) model
presented by Prugniel \& Simien (1997) (equation~\ref{r2_app} in this
paper) may be useful, if not even more appropriate, for such studies.

The new density--potential pair should also prove fruitful for
simulations of bulges in disk galaxies.  To date, such studies have
usually been performed using models which are inappropriate for
describing the majority of bulges.  This situation has arisen from
improved observations which have shown bulges typically have S\'ersic
indices $n$ ranging from 0.5 to 3, rather than 4 (e.g., Andredakis \&
Sanders 1994; de Jong 1996; Graham 2001; MacArthur et al.\ 2003). 
%

It is hoped that the models presented here will additionally enable
one to perform a number of simulations testing various observational
results and theoretical problems associated with the influence of
SMBHs on central density cusps and the connection with the global
stellar structure.  The influence of SMBHs on the stellar dynamics
(e.g., Baes, Dejonghe \& Buyle 2005) can also be explored as a
function of various physical parameters.
One way this can be achieved is by investigating the restrictions that
self--consistency requirements pose on galaxy density profiles (e.g.,
Terzi\'c 2003).
Other numerical simulations may investigate the effects of binary
SMBHs at the centers of galaxies (e.g., Graham 2004).  Gravitational
scattering (e.g., Makino \& Funato 2004), as well as resonant chaotic
phase mixing between the frequencies of the binary orbit and the
natural frequencies of stellar orbits, can cause a significant
redistribution of mass extending well beyond the radius of the binary
(Kandrup et al.\ 2003; see also Merritt \& Poon 2004).
One could also address the similarities and differences between the
nature and efficiency of chaotic phase mixing in the new model (both
time--independent and subjected to time--periodic pulsations) and
previous studies that have used somewhat more limited potentials
(e.g., Kandrup \& Mahon 1994, Merritt \& Valluri 1996; Siopis \&
Kandrup 2000, Terzi\'c \& Kandrup 2004).
In addition, one should be able to extend past work which has explored
the evolution of galaxies using models that had a restricted range of
outer galaxy structure (e.g., Merritt \& Fridman 1996; Poon \& Merritt
2002, 2004).
%
Furthermore, it is hoped one will explore {\it new} connections
between the outer galaxy structure, i.e.\ profile shape, and the
properties of the core, such as the inner cusp slope, core size, and
black hole mass.



\section{Summary}

Our application of the (2--parameter) Jaffe and Hernquist models to a
sample of luminosity--density profiles taken from real galaxies
reveals that these models are inadequate for describing the observed
range of stellar distributions.  The (3--parameter) Dehnen model does
better, providing a good match to the density profiles of large
early-type galaxies with S\'ersic indices around 4 and greater, but it 
fails to adequately match the deprojected light--profiles of galaxies
with low S\'ersic indices, which would encompass dwarf elliptical
galaxies and most bulges in disk galaxies (e.g., Balcells et al.\ 2003,
and references therein).
If one wishes to explore the hierarchical merging theory, in
particular the potential build up of large elliptical galaxies through
the collision of lesser elliptical galaxies, 
then none of the above models are good starting points.

These failures are not particularly surprising because all three of
these models were constructed to have outer density profiles that fall
off as $r^{-4}$ --- in order to match the decline in density of a
deprojected $R^{1/4}$ profile.  As such, they are highly useful but
nonetheless limited models.  The (3--parameter) density profile of
Prugniel \& Simien (1997), however, does not have the above
constraint, and consequently does much better, accurately matching the
density profiles of luminous elliptical galaxies with $n>4$ {\it and}
faint ellipticals with $n<4$, including exponential ($n=1$) profiles.
This model may even prove applicable to the density profiles of
$\Lambda$CDM--generated dark matter halos.

We have derived exact expressions for the gravitational potential and
force associated with the latter density profile.  The use of this
family of expressions for simulations of elliptical galaxies and
bulges in general, and explorations of how their structure may evolve
under various circumstances, should enable projects of a nature
previously prohibited by the former class of models.  Moreover, the
equations for the potential and force contain only one elementary
function and are otherwise analytical; they are therefore fast to
compute.

We have also developed what we believe to be the only density profile 
capable of simultaneously matching both the nuclear and global stellar
distribution in galaxies having partially depleted cores.  Application
of this (5--parameter) profile to NGC~3348 reveals a good fit (rms
$\sim$ 0.03 dex) over 3 orders of magnitude in radial range and 6 in
luminosity density.  Similar results are obtained from the core
galaxies NGC 4291 and NGC 2986.  Furthermore, the associated equations
for the potential and force are also derived here using only the
incomplete gamma function and analytical terms.  Expressions to derive
the enclosed mass, the spatial and projected velocity dispersion, and
the projected intensity profile are additionally provided.

\section{acknowledgments}

We are happy to thank Chris Hunter, Andi Burkert and Agris Kalnajs 
for their helpful comments and suggestions.  We also wish 
to thank the referee, Eric Emsellem, for his comments on the text. 
We are grateful to Nicola Caon for providing us with the
light--profiles for NGC~1379, 4374, 4458 \& 4552, to Peter Erwin for
kindly providing us with the light--profile for NGC~3348 and to
Massimo Stiavelli for supplying us with the light--profile for LGC~47.
This research was supported in part by NSF grant AST-0307351, 
Department of Energy grant G1A62056
and by NASA grant HST-AR-09927.01-A
from the Space Telescope Science Institute, which is operated by the
Association of Universities for Research in Astronomy, Inc., under
NASA contract NAS5-26555.

\newpage
\section{APPENDIX A: Mass, circular velocity, and velocity dispersion}
\subsection{Prugniel--Simien density profile}

In this section we provide some additional helpful expressions related
to the density profile $\rho(r)$ given in Prugniel \& Simien (1997,
their equation~B6), and given here by equation~(\ref{r2_app}).

Assuming spherical symmetry, the enclosed mass, $M(r)$, is simply 
\begin{equation}
M(r)  = 4\pi \int\limits_0^r \rho({\bar r}) {\bar r}^2 ~ d{\bar r}. 
\end{equation}
Using the change of variable ${\bar Z}=b({\bar r}/R_{\rm e})^{1/n}$, such that
$d{\bar r} = R_{\rm e} n b^{-n} {\bar Z}^{n-1} d{\bar Z}$, one obtains
\begin{equation} \label{M_PS}
M(r)  = {4\pi \rho_0 {R_{\rm e}}^3n} {b^{n(p-3)}}
\gamma\left(n(3-p),Z\right), 
\end{equation}
where $Z=b(r/R_{\rm e})^{1/n}$ and $\gamma(a,x)$ is the incomplete gamma
function given in equation~(\ref{eqgam}).
The total mass is obtained by replacing $\gamma\left(n(3-p),Z\right)$
with $\Gamma\left(n(3-p)\right)$. 

The circular velocity is given by 
\begin{equation} \label{vcirc}
v_{\rm circ}(r) = \sqrt{ \frac{GM(r)}{r}}. 
\end{equation}

Assuming isotropy, the spatial velocity dispersion is given in Binney
(1980) as
\begin{eqnarray} \label{sigma_PS}
{\sigma_{\rm s}}^2 (r)& = & {G \over {\rho(r)}} \int_r^{\infty} \rho({\bar r}) 
{{M({\bar r})} \over {\bar r}^2} d{\bar r} \nonumber \\
& = & {{4 \pi G {\rho_0}^2 {R_{\rm e}}^2 n^2 b^{2n(p-1)}} \over{\rho(r)}}
\int_Z^{\infty} {\bar Z}^{-n(p+1)-1} {\rm e}^{-{\bar Z}} \gamma \left(n(3-p),
{\bar Z} \right) d{\bar Z}. 
\end{eqnarray}
This expression is easily computed numerically after making the
substitution ${\bar Z}=(Z/\cos \theta)$, such that 
$d{\bar Z}/d\theta = Z\sin\theta/\cos^2\theta$, giving 
\begin{equation}
{\sigma_{\rm s}}^2 (r) = 
{4 \pi G (\rho_0 R_{\rm e} n b^{n(p-1)})^2 Z^{-n(p+1)} 
\over{\rho(r)} }
\int_0^{\pi/2} \tan\theta (\cos\theta)^{n(p+1)} {\rm e}^{-Z/\cos\theta}
\gamma \left(n(3-p), \frac{Z}{\cos\theta} \right) d\theta.
\end{equation}

The projected, line--of--sight velocity dispersion as given by 
equation (B12) in Prugniel \& Simien (1997) is
\begin{eqnarray} \label{sigma_los_PS}
{\sigma_{\rm p}}^2(R) & = & \frac{2G}{I(R)M_{\rm tot}} \int\limits_R^{\infty} 
  \frac{\sqrt{{\bar r}^2-R^2}}{{\bar r}^2} \rho({\bar r}) 
   M({\bar r}) ~ d{\bar r} \nonumber \\
& = & {{8 \pi G {\rho_0}^2 {R_{\rm e}}^3 n^2 b^{n(2p-3)}} \over {I(R) M_{\rm tot}}}
\int_Z^{\infty} 
\sqrt{{\bar Z}^{2n} - Z^{2n}} \hskip5pt {\bar Z}^{-n(p+1)-1} e^{-{\bar Z}} 
\gamma \left(n(3-p), {\bar Z} \right) d{\bar Z} \nonumber \\
& = & {{8 \pi G {\rho_0}^2 {R_{\rm e}}^3 n^2 b^{n(2p-3)}} \over {I(R) M_{\rm tot}}}
Z^{-np} \int_0^{\pi/2} 
\sqrt{1 - (\cos\theta)^{2n}}
\tan\theta (\cos\theta)^{np} {\rm e}^{-Z/\cos\theta}
\gamma \left(n(3-p), \frac{Z}{\cos\theta} \right) d\theta, 
\end{eqnarray}
with $Z$ now equal to $b(R/R_{\rm e})^{1/n}$. 

\begin{figure*}
\begin{center}
\begin{minipage}{168mm}
\includegraphics[width=168mm]{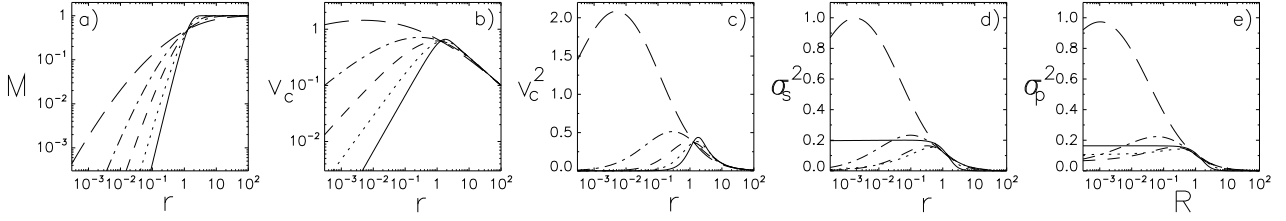}
\vskip30pt
\caption{Core--less galaxies.
Panel a) mass (equation~\ref{M_PS}),
b) circular velocity (equation~\ref{vcirc}), 
c) circular velocity squared, 
d) spatial velocity dispersion squared (equation~\ref{sigma_PS}), and
e) the square of the projected, line--of--sight velocity dispersion 
(equation~\ref{sigma_los_PS}), 
for varying values of the profile shape
$n$: $n=0.5$ (solid lines), $n=1$ (dotted), $n=2$ (dashed),
$n=4$ (dash-dot), $n=10$ (double--dash).
We have set $R_{\rm e}$=1, $M_{\rm tot}=1$, and $G=1$. 
The square of each velocity term is thus effectively normalised to 
$GM_{\rm tot}/R_{\rm e}$. 
Spherical symmetry and isotropy have been assumed.
}
\label{fig71}
\end{minipage}
\end{center}
\end{figure*}

The mass profiles, circular velocity profiles, spatial and projected
velocity dispersion profiles are shown in Fig.~\ref{fig71} for a range
of profile shapes.  They agree closely with those obtained from 
the exact deprojection of S\'ersic's $R^{1/n}$ model (Ciotti 1991; 
Simonneau \& Prada 2004).  In order to show the behavior of 
$v_{\rm circ}$ near the break radius $r_b$, and to allow comparisons
of $(v_{\rm circ})^2$ with $\sigma^2$, we present both $v_{\rm circ}$
on a logarithmic scale and $(v_{\rm circ})^2$ on a linear scale.

%
%

\subsection{Our density profile}

Below are the equivalent expressions relevant to our adaptation of
Prugniel \& Simien's density profile for galaxies with power--law cores.

The enclosed mass $M(r)$ is 
\begin{equation} \label{M_our}
M(r) = 4\pi \int\limits_0^r \rho({\bar r}) {\bar r}^2 ~ d{\bar r}
     = 4\pi \int\limits_0^{r_{\rm min}} \rho_1({\bar r}) {\bar r}^2 ~ d{\bar r}
      +4\pi \int\limits_{r_{\rm min}}^r \rho_2({\bar r}) {\bar r}^2 ~ d{\bar r} 
     \equiv M_1(r) + M_2(r)
\end{equation}
where $\rho_1({\bar r})$ and $\rho_2({\bar r})$ are given in 
equation~(\ref{rhos}), $r_{\rm min}$ is the minimum value of $r$ or $r_b$, 
and 
\begin{equation}
M_1(r) = 4\pi J_1(r),
\end{equation}
\begin{equation}
M_2(r) = 4\pi L_2(r) , 
\end{equation}
where $J_1(r)$ and $L_2(r)$ are defined in equations~(\ref{J_1}) and (\ref{L_2}),
respectively.
The total mass is obtained by setting the second incomplete gamma
function in the expression for $L_2(r)$ (equation~\ref{L_2}) equal to zero. 
Assuming spherical symmetry, the circular velocity is given by 
equation~(\ref{vcirc}).

Again assuming isotropy, the spatial velocity dispersion is given by
\begin{equation} \label{sigma_our}
{\sigma_{\rm s}}^2 (r) \equiv K_1(r) + K_2(r) + K_3(r),
\end{equation}
where
\begin{eqnarray}
K_1(r) 
& = & {G \over {\rho_1(r)}} 
\int_r^{r_{\rm max}} \rho_1({\bar r}) {{M_1({\bar r})} \over {\bar r}^2} d{\bar r} 
\nonumber \\
& = & {{4 \pi G {\rho_b}^2 {r_b}^{2 \gamma}} \over {\rho_1(r)}} 
\left\{\begin{array}{ll}
{1 \over {3-\gamma}} \ln {r_{\rm max} \over r} & \mbox{if $\gamma = 1$} \\
{1 \over {2(3-\gamma)(1-\gamma)}} 
\left({r_{\rm max}}^{2(1-\gamma)}-r^{2(1-\gamma)}\right) 
  & \mbox{if $\gamma \ne 1$ and $< 3$} 
\end{array} \right.,
\end{eqnarray}
\begin{eqnarray}
K_2(r) 
& = & {G \over {\rho_2(r)}} 
M_1(r_b) \int_{r_{\rm max}}^{\infty} {{\rho_2({\bar r})} \over {\bar r}^2} d{\bar r} 
\nonumber \\
& = & {{G M_1(r_b) \rho_b {\bar \rho} n b^{n(p+1)}} \over {R_{\rm e}\rho_2(r)}} 
\int_{Z_{\rm max}}^{\infty} {\bar Z}^{-n(p+1)-1} e^{-{\bar Z}} d{\bar Z}, 
\end{eqnarray}
\begin{eqnarray}
K_3 (r) 
& = & {G \over {\rho_2(r)}} 
\int_{r_{\rm max}}^{\infty} \rho_2({\bar r}) {{M_2({\bar r})} \over {\bar r}^2} d{\bar r}
\nonumber \\
& = & {{4 \pi G {\rho_b}^2 {\bar \rho}^2 {R_{\rm e}}^2 n^2 b^{2n(p-1)}} \over {\rho_2(r)}} \nonumber \\
& \times & \int_{Z_{\rm max}}^{\infty} {\bar Z}^{-n(p+1)-1} e^{-{\bar Z}} 
\left[
   \Gamma \left(n(3-p),Z_b\right)
  -\Gamma \left(n(3-p),{\bar Z}\right)
  \right] d{\bar Z}, 
\end{eqnarray}
with $r_{\rm max}=$ max$(r,r_b)$, 
$Z=b(r/R_{\rm e})^{1/n}$, 
$Z_b=b(r_b/R_{\rm e})^{1/n}$ 
and $Z_{\rm max}=$ max$(Z,Z_b)$.
In passing, we note that the 
term $-n(p+1)$ appearing in the expression for $K_2(r)$ equals
$0.6097-2n-0.05563/n$, and is thus negative for values of $n$ greater
than about 0.2. For this reason we did not express $K_2(r)$ in terms of
the incomplete gamma function. 
Following what was done with equation~(\ref{sigma_PS}), 
to help evaluate $K_2(r)$ and $K_3(r)$ numerically 
one may apply the change of variable ${\bar Z} = Z_{\rm max}/\cos\theta$, to give
\begin{equation}
K_2(r) = {{G M_1(r_b) \rho_b {\bar \rho} n b^{n(p+1)}} \over {R_{\rm e}\rho_2(r)}}
(Z_{\rm max})^{-n(p+1)}
\int_0^{\pi/2} \tan\theta (\cos\theta)^{n(p+1)} {\rm e}^{-Z_{\rm max}/\cos\theta} d\theta,
\end{equation}
\begin{eqnarray}
K_3(r) & = & {{4 \pi G {\rho_b}^2 {\bar \rho}^2 {R_{\rm e}}^2 n^2 b^{2n(p-1)}} \over {\rho_2(r)}}
(Z_{\rm max})^{-n(p+1)} \nonumber \\
 & \times &
\int_0^{\pi/2} \tan\theta (\cos\theta)^{n(p+1)} {\rm e}^{-Z_{\rm max}/\cos\theta} 
\left[
   \Gamma \left(n(3-p),Z_b\right)
  -\Gamma \left(n(3-p),Z_{\rm max}/\cos\theta \right)
  \right]
d\theta.
\end{eqnarray}
Alternatively, one may use the 
change of variable ${\bar r} = 1/s$ to transform the $K_2(r)$ term into 
the expression 
\begin{equation}
K_2(r) = {G \over {\rho_2(r)}} M_1(r_b) \int_0^{1/r_{\rm max}}
\rho_2(1/s) ds. 
\end{equation}

The projected, line--of--sight velocity dispersion is found to be
\begin{equation} \label{sigma_los_our}
{\sigma_{\rm p}}^2(R) \equiv S_1(R) + S_2(R) + S_3(R),
\end{equation}
where 
\begin{eqnarray} \label{S_1}
S_1(R) & = & \frac{2G}{I(R)M_{\rm tot}} 
\int\limits_R^{R_{\rm max}} \frac{\sqrt{{\bar r}^2-R^2}}{{\bar r}^2} 
\rho_1({\bar r}) M_1({\bar r}) ~ d{\bar r}  \nonumber \\
& = & \frac{8 \pi G {\rho_b}^2 {r_b}^{2 \gamma} {R_{\rm e}}^{3-2\gamma} 
n b^{n(2\gamma-3)}} {I(R)M_{\rm tot}(3-\gamma)} 
\int^{Z_{\rm max}}_{Z}
\sqrt{{\bar Z}^{2n} - Z^{2n}} {\bar Z}^{2n(1-\gamma)-1} d{\bar Z}
\hskip10pt  \mbox{if $\gamma < 3$},
\end{eqnarray}
\begin{eqnarray} \label{S_2}
S_2(R) & = & \frac{2G}{I(R)M_{\rm tot}} 
M_1({r_b}) \int\limits_{R_{\rm max}}^{\infty} 
\frac{\sqrt{{\bar r}^2-R^2}}{{\bar r}^2} \rho_2({\bar r}) ~ d{\bar r} 
\nonumber \\
& = & \frac{2G}{I(R)M_{\rm tot}} M_1(r_b) n b^{np} \rho_b {\bar \rho}
\int_{Z_{\rm max}}^{\infty} 
\sqrt{{\bar Z}^{2n} - Z^{2n}} {\bar Z}^{-n(p+1)-1} e^{-{\bar Z}} d{\bar Z},
\end{eqnarray}
\begin{eqnarray} \label{S_3}
S_3(R) & = & \frac{2G}{I(R)M_{\rm tot}} 
\int\limits_{R_{\rm max}}^{\infty} \frac{\sqrt{{\bar r}^2-R^2}}{{\bar r}^2} 
\rho_2({\bar r}) M_2({\bar r}) ~ d{\bar r} \nonumber \\
& = & {{8 \pi G {\rho_b}^2 {\bar \rho}^2 {R_{\rm e}}^3 n^2 b^{n(2p-3)}}
\over {I(R) M_{\rm tot}}} \nonumber \\
& \times & \int_{Z_{\rm max}}^{\infty} 
\sqrt{{\bar Z}^{2n} - Z^{2n}} {\bar Z}^{-n(p+1)-1} e^{-{\bar Z}} 
\left[
   \Gamma \left(n(3-p),Z_b\right)
  -\Gamma \left(n(3-p),{\bar Z}\right)
  \right] d{\bar Z}, 
\end{eqnarray}
with $R_b = r_b$, 
$R_{\rm max}=$max$(R,R_b)$,
$Z=b(R/R_{\rm e})^{1/n}$ and 
$Z_b=b(R_b/R_{\rm e})^{1/n}$. 
Equation~\ref{S_2} and \ref{S_3} can be integrated numerically 
using the change of variable ${\bar Z} = Z_{\rm max}/\cos\theta$.

\begin{figure*}
\begin{center}
\begin{minipage}{168mm}
\includegraphics[width=168mm]{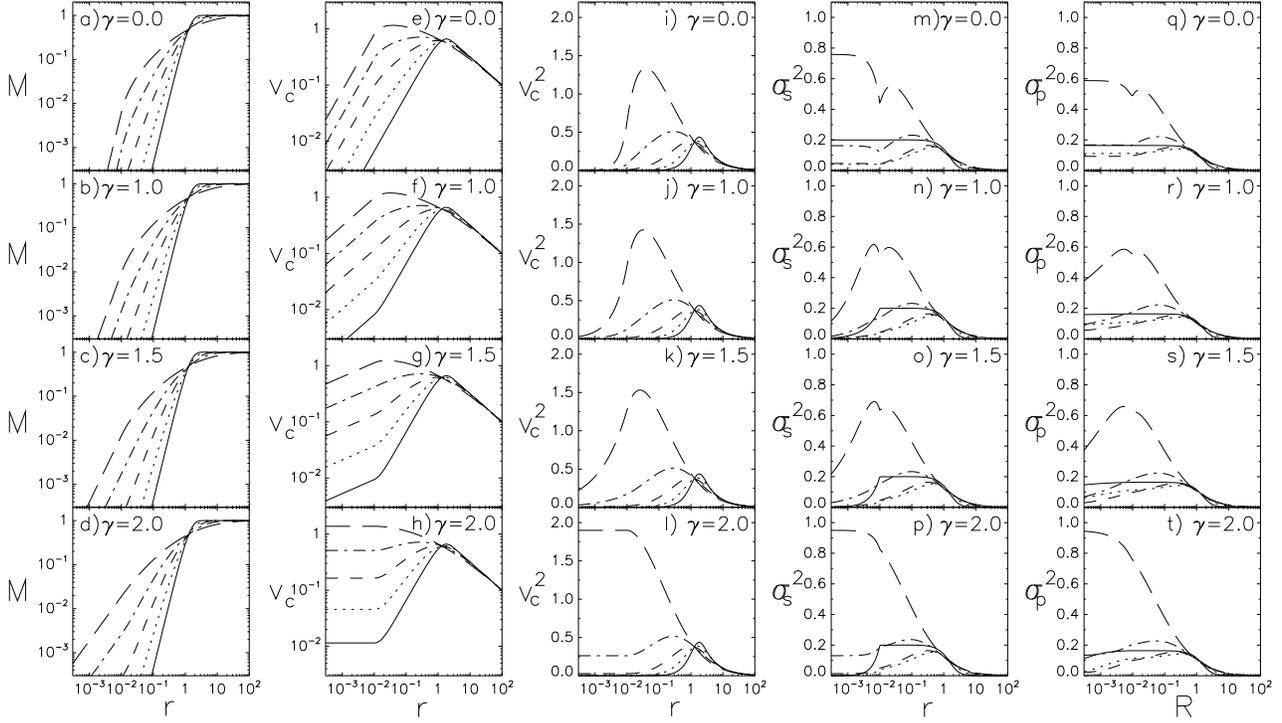}
\vskip30pt
\caption{Core galaxies.
Panels a--d) show the normalised mass (equation~\ref{M_our}),
e--h) circular velocity (equation~\ref{vcirc}), 
i--l) circular velocity squared, 
m--p) spatial velocity dispersion squared (equation~\ref{sigma_our}) and 
q--t) the square of the projected, line--of--sight velocity dispersion 
(equation~\ref{sigma_los_our}), for varying values of the outer profile shape 
$n$: $n=0.5$ (solid lines), $n=1$ (dotted), $n=2$ (dashed),
$n=4$ (dash-dot), $n=10$ (double--dash); and varying central cusp 
slope $\gamma$. 
We have set $R_{\rm e}=1$, $r_b=0.01$, $M_{\rm tot}=1$, and $G=1$. 
The square of each velocity term is thus effectively normalised to 
$GM_{\rm tot}/R_{\rm e}$. 
Spherical symmetry and isotropy have been assumed.
To show the behavior of $v_{\rm circ}$ near $r_b$, and to enable easy 
comparison of $(v_{\rm circ})^2$ with $\sigma_{\rm s}^2$ and $\sigma_{\rm p}^2$,
we present $v_{\rm circ}$ on a logarithmic scale and $(v_{\rm circ})^2$ 
on a linear scale.} 
\label{fig72}
\end{minipage}
\end{center}
\end{figure*}

The mass profiles, circular velocity profiles, spatial and projected
velocity dispersion profiles are shown in Fig.~\ref{fig72} for a range
of profile shapes.
The intriguing ``dip'' in some of the velocity dispersion profiles,
most notably the high $n$ profiles with small $\gamma$, is perhaps worth
commenting on.  It arises from the use of a sharp transition in the
density profile (equation~\ref{cS2_2}) when the inner and outer slope
(on either side of $r_b$) are markedly different.  Application of a
model with an extended transition region quenches this dip.  The use
of equation~\ref{cS2_1} with decreasing values of $\alpha$, i.e. an
increasingly broader transition region, steadily reduces and eliminates the
dip.  While profiles with $\alpha=10$ largely reproduce the velocity
dispersions shown, density profiles with $\alpha=1$ have smooth
velocity dispersion profiles.  
In any case, an inspection of the ``pseudo'' potential
(equation~\ref{pseudo}) reveals that such a velocity structure does
not give rise to an unstable situation for particle orbits.

Equations pertaining to 
Prugniel \& Simien's (1997) density profile are obtained by setting 
$r_b=0$ and $\rho_b {\bar \rho} = \rho_0$.  Indeed, upon making
these substitutions in the derivation of equations for enclosed mass
and velocity dispersion above, $r_{\rm min} = 0$, $r_{\rm max}=r$, $R_{\rm max}=R$,
$Z_{\rm max}=Z$ and the terms $K_1(r)$, $K_2(r)$, $S_1(r)$ and $S_2(r)$
vanish.  Equations (\ref{M_PS}), (\ref{sigma_PS}) 
and (\ref{sigma_los_PS}) are then easily recovered.


\newpage
\section{APPENDIX B: Derivation of the potential, force and intensity profiles}

\subsection{Potential}

If the expression for the density (equation~\ref{cS2_2}) is rewritten
as $\rho(r)=\rho_1(r) + \rho_2(r)$ (as was done in
subsection~\ref{secSB}, equation~\ref{rhos}), then the potential
(equation~\ref{BT}) can be written as
\begin{equation} \label{phi_1}
\Phi(r) =
{-4 \pi G}
\left\{
  \begin{array}{ll}
{1 \over r} \int\limits^r_0 \rho_1({\bar r}) {\bar r}^2 ~ d{\bar r} +
\int\limits^{r_b}_r \rho_1({\bar r}) {\bar r} ~ d{\bar r} +
\int\limits^{\infty}_{r_b} \rho_2({\bar r}) {\bar r} ~ d{\bar r}
& \mbox{if $r \le r_b$}, \\
{1 \over r} \int\limits^{r_b}_0 \rho_1({\bar r}) {\bar r}^2 ~ d{\bar r} +
{1 \over r} \int\limits^r_{r_b} \rho_2({\bar r}) {\bar r}^2 ~ d{\bar r} +
\int\limits^{\infty}_r \rho_2({\bar r}) {\bar r} ~ d{\bar r}
& \mbox{if $r > r_b$},
  \end{array} \right..
\end{equation}
\begin{equation}
\Phi(r) =
{-4 \pi G}
\left\{
  \begin{array}{ll}
{1 \over r} J_1(r) + J_2(r) + L_1(r_b)
& \mbox{if $r \le r_b$}, \\
{1 \over r} J_1(r_b) + {1 \over r} L_2(r) + L_1(r)
& \mbox{if $r > r_b$},
  \end{array} \right.,
\end{equation}
The derivation of integrals $J_1(r)$ and $J_2(r)$, which only involve a 
power-law integrand associated with $\rho_1(r)$, is trivial, while the 
integrals which involve $\rho_2(r)$ require a change of variable
${\bar Z} \equiv b ({\bar r}/R_{\rm e})^{1/n}$, so that ${\bar r}=
R_{\rm e}({\bar Z}/b)^n$ and 
$d{\bar r}=R_{\rm e}n{\bar Z}^{n-1}b^{-n} d{\bar Z}$: 
\begin{eqnarray} \label{Ls}
L_1(r) & \equiv &
\int\limits^{\infty}_r \rho_2({\bar r}) {\bar r} ~ d{\bar r}
= \rho_b {\bar \rho}
\int\limits^{\infty}_r \left({{\bar r} \over R_{\rm e}}\right)^{-p}
{\rm e}^{-b\left({\bar r}/R_{\rm e}\right)^{1/n}} {\bar r} ~ d{\bar r} \nonumber \\
& = & \rho_b {\bar \rho} R_{\rm e}
\int\limits^{\infty}_r \left({{\bar r} \over R_{\rm e}}\right)^{1-p}
{\rm e}^{-b\left({\bar r}/R_{\rm e}\right)^{1/n}} ~ d{\bar r} \nonumber \\
& = & \rho_b {\bar \rho} {R_{\rm e}}^2 n b^{n(p-2)}
\int\limits^{\infty}_{b\left(r/R_{\rm e}\right)^{1/n}}
{\bar Z}^{n(2-p)-1} {\rm e}^{-{\bar Z}} ~ d{\bar Z} \nonumber \\
& = & \rho_b {\bar \rho} {R_{\rm e}}^2 n b^{n(p-2)}
  \Gamma\left(n(2-p),b\left({r \over R_{\rm e}}\right)^{1/n}\right) \\
L_2(r) & \equiv &
\int\limits_{r_b}^r \rho_2({\bar r}) {\bar r}^2 ~ d{\bar r}
= \rho_b {\bar \rho}
\int\limits_{r_b}^r \left({{\bar r} \over R_{\rm e}}\right)^{-p}
{\rm e}^{-b\left({\bar r}/R_{\rm e}\right)^{1/n}} {\bar r}^2 ~ d{\bar r} \nonumber \\
& = & \rho_b {\bar \rho} {R_{\rm e}}^2
\int\limits_{r_b}^r \left({{\bar r} \over R_{\rm e}}\right)^{2-p}
{\rm e}^{-b\left({\bar r}/R_{\rm e}\right)^{1/n}} ~ d{\bar r} \nonumber \\
& = & \rho_b {\bar \rho} {R_{\rm e}}^3 n b^{n(p-3)}
\int\limits^{b\left(r/R_{\rm e}\right)^{1/n}}_{b\left(r_b/R_{\rm e}\right)^{1/n}}
{\bar Z}^{n(3-p)-1} {\rm e}^{-{\bar Z}} ~ d{\bar Z} \nonumber \\
& = & \rho_b {\bar \rho} {R_{\rm e}}^3 n b^{n(p-3)}
\left[
  \Gamma \left(n(3-p),b\left({r_b, \over R_{\rm e}}\right)^{1/n}\right)
 -\Gamma \left(n(3-p),b\left({r \over R_{\rm e}}\right)^{1/n}\right)
\right].
\end{eqnarray}

\subsection{Force}

Computing the radial force requires differentiation of the potential 
(\ref{pot_cS2}) with respect to the radial coordinate $r$.  When 
$r \le r_b$, the differentiation is trivial after realising that 
the last term is constant.  For $r > r_b$, we use the relation 
given in Abramowitz \& Stegun (1974, their equation~6.5.25): 
\begin{equation} \label{gamma_diff}
{\partial \Gamma(a,x) \over \partial x} = -x^{a-1} {\rm e}^{-x}, 
\end{equation}
%
%
%
%
%
and we note the following cancellation of terms simplifies matters
\begin{equation} \label{cancel}
{1 \over r} {{d L_2(r)} \over {d r}} = - {{d L_1(r)} \over {d r}} =
\rho_b {\bar \rho} R_{\rm e} \left({r\over R_{\rm e}}\right)^{1-p} 
{\rm e}^{-b\left(r/R_{\rm e}\right)^{1/n}}.
\end{equation}


\subsection{Projected Intensity}

Substituting $\rho_1(r)$ from equation~(\ref{rhos}) into
equation~(\ref{rho2I}) yields
\begin{equation} \label{A_I1}
I_1(R) = {2 \over \Upsilon} \rho_b {r_b}^{\gamma}
\int\limits^{r_b}_R {r^{1-\gamma}\over{\sqrt{r^2-R^2}}} dr.
\end{equation}
By using the change of variable $y=r^2$, 
it is trivial to compute the expression for the 
special case when $\gamma=0$.  The solution is given 
in the first line of equation~(\ref{I1_cS2}).  

When $\gamma=1$, the change of variable $r=R/\cos\theta$, such that 
 $dr/d\theta = R\sin\theta/\cos^2\theta$, gives
\begin{eqnarray}
I_1(R) & = & {2 \over \Upsilon} \rho_b {r_b}^{\gamma}
\int\limits^{r_b}_R {1 \over{\sqrt{r^2-R^2}}} dr 
 = {2 \over \Upsilon} \rho_b {r_b}^{\gamma}
\int\limits_0^{\cos^{-1}(R/r_b)} {1 \over \cos\theta} d\theta \nonumber \\
 & = & {2 \over \Upsilon} \rho_b {r_b}^{\gamma}
\left.
\ln \left({1 \over \cos\theta} + {\sin\theta \over \cos\theta} \right)
   \right|_0^{\cos^{-1}(R/r_b)}
 = {2 \over \Upsilon} \rho_b {r_b}^{\gamma}
\left.
\ln \left({1+\sqrt{1-\cos^2\theta} \over \cos\theta} \right)
   \right|_0^{\cos^{-1}(R/r_b)} \nonumber \\
 & = & {2 \over \Upsilon} \rho_b {r_b}^{\gamma}
\ln \left( { 1+\sqrt{1-R^2/{r_b}^2} \over {R/r_b} } \right) 
 = {2 \over \Upsilon} \rho_b {r_b}^{\gamma}
\ln \left( { r_b+\sqrt{r_b^2-R^2} \over {R} } \right). 
\end{eqnarray}

For other values of the cusp slope $\gamma$, further transformation of 
equation~(\ref{A_I1}) is required.  Using the change of variable 
$y \equiv 1 - R^2/r^2$, equation~(\ref{A_I1}) becomes
\begin{eqnarray} \label{A_I3}
I_1(R) & = & {2 \over \Upsilon} \rho_b {r_b}^{\gamma} \left[
{1 \over 2} R^{1-\gamma}
\int\limits^{1-R^2/{r_b}^2}_{0} y^{-1/2}
(1-y)^{(\gamma-3)/2} dy \right] \nonumber \\
 & = & {\rho_b {r_b}^{\gamma} \over \Upsilon}
R^{1-\gamma}
B_{1-R^2/{r_b}^2} \left({1 \over 2}, {{\gamma-1}\over{2}}\right).
\end{eqnarray}

When $\gamma=2$, the incomplete beta function (equation~\ref{BETA})
can be simplified through the change of variable $u=t^2$, to give
\begin{eqnarray}
I_1(R) & = & {2 \over \Upsilon} \rho_b {r_b}^{\gamma} \left[
R^{-1} \int\limits^{\sqrt{1-R^2/{r_b}^2}}_{0}
{1 \over \sqrt{1-t^2}} dt \right] 
 = {2\rho_b {r_b}^{\gamma} \over \Upsilon} 
R^{-1} \sin^{-1}\left( \sqrt{1-R^2/r_b^2} \right). 
\end{eqnarray}

\label{lastpage}
\end{document}